%% file: main.tex
\documentclass[compsoc,conference,a4paper,10pt,times]{IEEEtran}

\usepackage{cite}
\usepackage[dvipsnames,x11names]{xcolor}
\usepackage{graphicx}
\usepackage{tikzpeople}[4]
\usepackage[notext,not1,notextcomp]{stix2}
\usepackage{pifont}
\usepackage{fontawesome5}
\usepackage{placeins}
\usepackage{scalefnt}
\usepackage{circledsteps}
\usepackage[edges]{forest}
\usepackage{tabularx}
\usepackage{adjustbox}
\usepackage{keyvaltable}
\usepackage{threeparttable}
\usepackage{soul}
\usepackage{xstring}
\usepackage{listings}
\usepackage[hyphens]{url}
\usepackage[colorlinks=true,urlcolor=black]{hyperref}
\usepackage[hyphenbreaks]{breakurl}
\usepackage[textsize=footnotesize]{todonotes}

\usetikzlibrary{patterns}
\usetikzlibrary{arrows.meta,calc}

\ifdefined\AnonymousReview
  \setuptodonotes{disable}
\fi

\ifdefined\AnonymousReview

\fi

\definecolor{ACMRed}{rgb}{.992,.106,.078}
\definecolor{ACMOrange}{rgb}{.988,.573,0}
\definecolor{ACMPurple}{rgb}{.396,.004,.42}
\definecolor{ACMBlue}{rgb}{.004,0.51,.675}
\definecolor{ACMDarkBlue}{rgb}{.035,.208,.478}

\colorlet{ColorA}{Black}
\colorlet{ColorB}{RubineRed!20!black}
\colorlet{ColorC}{Orange!20!black}
\colorlet{ColorD}{ACMPurple!40!black}
\colorlet{ColorE}{Sepia!40!black}
\colorlet{ColorF}{ACMDarkBlue!40!black}
\colorlet{ColorG}{ForestGreen!20!black}

\colorlet{FillA}{Black!15}
\colorlet{FillB}{RubineRed!15}
\colorlet{FillC}{Orange!15}
\colorlet{FillD}{ACMPurple!15}
\colorlet{FillE}{Sepia!15}
\colorlet{FillF}{ACMDarkBlue!15}
\colorlet{FillG}{ForestGreen!15}

\colorlet{OutlineA}{Black!50}
\colorlet{OutlineB}{RubineRed!50}
\colorlet{OutlineC}{Orange!50}
\colorlet{OutlineD}{ACMPurple!50}
\colorlet{OutlineE}{Sepia!50}
\colorlet{OutlineF}{ACMDarkBlue!50}
\colorlet{OutlineG}{ForestGreen!50}

\pgfkeys{/csteps/inner color=ACMDarkBlue!50}
\pgfkeys{/csteps/outer color=ACMDarkBlue!20}

\begin{document}

\setlength{\marginparwidth}{15mm}

\makeatletter
\@ifdefinable\anon{%
  \newcommand{\anon}[2][ANONYMIZED]{%
    \@ifundefined{AnonymousReview}{#2}{#1}%
  }%
}
\makeatother

\newcommand{\citeyear}[1]{%
  \StrBefore{#1}{:}[\TmpCiteYearBefore]%
  \StrRight{\TmpCiteYearBefore}{4}[\TmpCiteYearRight]%
  \TmpCiteYearRight%
}

\newcommand{\citeauthor}[1]{%
  \StrBefore{#1}{:}[\TmpCiteAuthorBefore]%
  \StrGobbleRight{\TmpCiteAuthorBefore}{4}[\TmpCiteAuthorGobbleRight]%
  \ifstrequal{#1}{Zhang2021:ThreeDecadesDeception}{Zhang and Thing}{%
    \ifstrequal{#1}{Araujo2020:ImprovingCybersecurityHygiene}{Araujo and Taylor}{%
      \ifstrequal{#1}{Almeshekah2014:PlanningIntegratingDeception}{Almeshekah and Spafford}{%
        \ifstrequal{#1}{Spitzner2003:HoneypotsCatchingInsider}{Spitzner}{%
          \ifstrequal{#1}{Kern2024:InjectingSharedLibraries}{Kern}{%
            \TmpCiteAuthorGobbleRight~et~al.}}}}}%
}

\title{Application Layer Cyber Deception without Developer Interaction}

\ifdefined\AnonymousReview
  \author{\IEEEauthorblockN{Anonymous Author(s)}
    \IEEEauthorblockA{\ignorespaces}
    \IEEEauthorblockA{\ignorespaces}
  }
\else
  \author{%
    \IEEEauthorblockN{%
      \href{https://orcid.org/0000-0002-6820-4953}{Mario Kahlhofer}}
    \IEEEauthorblockA{%
      Dynatrace Research}
    \IEEEauthorblockA{%
      mario.kahlhofer@dynatrace.com
    }
    \and
    \IEEEauthorblockN{%
      \href{https://orcid.org/0000-0003-2821-2489}{Stefan Rass}}
    \IEEEauthorblockA{%
      Johannes Kepler University Linz}
    \IEEEauthorblockA{%
      stefan.rass@jku.at
    }
  }
\fi

\lstset{%
  tabsize=2,
  frame=single,
  emptylines=1,
  captionpos=b,
  aboveskip=12pt,
  belowskip=6pt,
  abovecaptionskip=6pt,
  showspaces=false,
  keepspaces=true,
  columns=fullflexible,
  showstringspaces=false,
  escapeinside={(*@}{@*)},
  basicstyle=\linespread{0.9}\footnotesize\ttfamily
}

\input{./includes/circles.tex}
\input{./includes/marks.tex}
\input{./includes/cites.tex}

\newcommand{\tildecentered}{\raisebox{0.5ex}{\texttildelow}}
\newcommand{\numMethods}{19}

\maketitle

\begin{abstract}
  Cyber deception techniques that are tightly intertwined with applications
  pose significant technical challenges in production systems.
  Security measures are usually the responsibility of a system operator, but
  they are typically limited to accessing built software artifacts, not their source code.
  This limitation makes it particularly challenging to deploy
  cyber deception techniques at application runtime
  and without full control over the software development lifecycle.
  This work reviews \numMethods{}~technical methods
  to accomplish this and evaluates them 
  based on technical, topological, operational, and efficacy properties.
  We find some novel techniques beyond honeypots and reverse proxies
  that seem to have received little research interest
  despite their promise for cyber deception. 
  We believe that overcoming these technical challenges can drive the adoption
  of more dynamic and personalized cyber deception techniques,
  tailored to specific classes of applications.
\end{abstract}


\input{./content/01-introduction.tex}
\input{./content/02-problem.tex}
\input{./content/03-literature.tex}
\input{./content/04-properties.tex}
\input{./content/05-methods.tex}
\input{./content/06-discussion.tex}
\input{./content/07-conclusion.tex}

\ifdefined\AnonymousReview\else
  \section*{Acknowledgments}
  We thank Anis and Markus for reviewing the methods, their categorization, and terminology,
  and the anonymous reviewers for their clear and useful feedback on our work.
\fi

\clearpage

\bibliographystyle{IEEEtranS}
{\small\sloppy\hfuzz=2pt\bibliography{abbrev,references}}

\clearpage

\appendices
\input{./content/08-appendix.tex}

\end{document}

%% file: includes/circles.tex
\newcommand{\CircleA}[1]{%
  \CircledParamOpts{inner color=ColorA,outer color=FillA,fill color=FillA}{1}{#1}%
}
\newcommand{\CircleB}[1]{%
  \CircledParamOpts{inner color=ColorB,outer color=FillB,fill color=FillB}{1}{#1}%
}
\newcommand{\CircleC}[1]{%
  \CircledParamOpts{inner color=ColorC,outer color=FillC,fill color=FillC}{1}{#1}%
}
\newcommand{\CircleD}[1]{%
  \CircledParamOpts{inner color=ColorD,outer color=FillD,fill color=FillD}{1}{#1}%
}
\newcommand{\CircleE}[1]{%
  \CircledParamOpts{inner color=ColorE,outer color=FillE,fill color=FillE}{1}{#1}%
}
\newcommand{\CircleF}[1]{%
  \CircledParamOpts{inner color=ColorF,outer color=FillF,fill color=FillF}{1}{#1}%
}
\newcommand{\CircleG}[1]{%
  \CircledParamOpts{inner color=ColorG,outer color=FillG,fill color=FillG}{1}{#1}%
}

\newcommand{\BCircleA}[1]{%
  \CircledParamOpts{inner color=ColorA,outer color=OutlineA,fill color=FillA}{1}{#1}%
}
\newcommand{\BCircleB}[1]{%
  \CircledParamOpts{inner color=ColorB,outer color=OutlineB,fill color=FillB}{1}{#1}%
}
\newcommand{\BCircleC}[1]{%
  \CircledParamOpts{inner color=ColorC,outer color=OutlineC,fill color=FillC}{1}{#1}%
}
\newcommand{\BCircleD}[1]{%
  \CircledParamOpts{inner color=ColorD,outer color=OutlineD,fill color=FillD}{1}{#1}%
}
\newcommand{\BCircleE}[1]{%
  \CircledParamOpts{inner color=ColorE,outer color=OutlineE,fill color=FillE}{1}{#1}%
}
\newcommand{\BCircleF}[1]{%
  \CircledParamOpts{inner color=ColorF,outer color=OutlineF,fill color=FillF}{1}{#1}%
}
\newcommand{\BCircleG}[1]{%
  \CircledParamOpts{inner color=ColorG,outer color=OutlineG,fill color=FillG}{1}{#1}%
}

\newcommand{\tDeploymentAgent}{{\setulcolor{OutlineA}\ul{deployment agent}}}
\newcommand{\tKubernetesOperator}{{\setulcolor{OutlineA}\ul{K8s operator}}}
\newcommand{\tSecondaryContainers}{{\setulcolor{OutlineA}\ul{secondary containers}}}
\newcommand{\tLifecycleHooks}{{\setulcolor{OutlineA}\ul{lifecycle hooks}}}
\newcommand{\tHoneypots}{{\setulcolor{OutlineB}\ul{honeypots}}}
\newcommand{\tReverseProxies}{{\setulcolor{OutlineC}\ul{reverse proxies}}}
\newcommand{\tFileModifications}{{\setulcolor{OutlineD}\ul{container file modifications}}}
\newcommand{\tRuntimeInstrumentation}{{\setulcolor{OutlineD}\ul{runtime instrumentation}}}
\newcommand{\tLdPreload}{{\setulcolor{OutlineD}\ul{\texttt{LD\_PRELOAD}}}}
\newcommand{\tPtrace}{{\setulcolor{OutlineD}\ul{\texttt{ptrace}}}}
\newcommand{\tEbpfMonitoring}{{\setulcolor{OutlineE}\ul{eBPF}}}
\newcommand{\tNetfilterRouting}{{\setulcolor{OutlineE}\ul{Netfilter}}}
\newcommand{\tEbpfRouting}{{\setulcolor{OutlineE}\ul{eBPF}}}
\newcommand{\tFilesystemImplementations}{{\setulcolor{OutlineE}\ul{file system implementation}}}
\newcommand{\tVolumeMounts}{{\setulcolor{OutlineF}\ul{volume mounts}}}
\newcommand{\tPolicyEngine}{{\setulcolor{OutlineF}\ul{policy engine}}}
\newcommand{\tCni}{{\setulcolor{OutlineF}\ul{CNI plugin}}}
\newcommand{\tDeceptionFramework}{{\setulcolor{OutlineG}\ul{deception software framework}}}
\newcommand{\tBuildProcess}{{\setulcolor{OutlineG}\ul{build process intervention}}}

\newcommand{\tsDeploymentAgent}{{\setulcolor{OutlineA}\ul{Deployment agent}}}
\newcommand{\tsKubernetesOperator}{{\setulcolor{OutlineA}\ul{K8s operator}}}
\newcommand{\tsSecondaryContainers}{{\setulcolor{OutlineA}\ul{Secondary containers}}}
\newcommand{\tsLifecycleHooks}{{\setulcolor{OutlineA}\ul{Lifecycle hooks}}}
\newcommand{\tsHoneypots}{{\setulcolor{OutlineB}\ul{Honeypots}}}
\newcommand{\tsReverseProxies}{{\setulcolor{OutlineC}\ul{Reverse proxies}}}
\newcommand{\tsFileModifications}{{\setulcolor{OutlineD}\ul{Container file modifications}}}
\newcommand{\tsRuntimeInstrumentation}{{\setulcolor{OutlineD}\ul{Runtime instrumentation}}}
\newcommand{\tsLdPreload}{{\setulcolor{OutlineD}\ul{\texttt{LD\_PRELOAD}}}}
\newcommand{\tsPtrace}{{\setulcolor{OutlineD}\ul{\texttt{ptrace}}}}
\newcommand{\tsEbpfMonitoring}{{\setulcolor{OutlineE}\ul{eBPF}}}
\newcommand{\tsNetfilterRouting}{{\setulcolor{OutlineE}\ul{Netfilter}}}
\newcommand{\tsEbpfRouting}{{\setulcolor{OutlineE}\ul{eBPF}}}
\newcommand{\tsFilesystemImplementations}{{\setulcolor{OutlineE}\ul{File system implementation}}}
\newcommand{\tsVolumeMounts}{{\setulcolor{OutlineF}\ul{Volume mounts}}}
\newcommand{\tsPolicyEngine}{{\setulcolor{OutlineF}\ul{Policy engine}}}
\newcommand{\tsCni}{{\setulcolor{OutlineF}\ul{CNI plugin}}}
\newcommand{\tsDeceptionFramework}{{\setulcolor{OutlineG}\ul{Deception software framework}}}
\newcommand{\tsBuildProcess}{{\setulcolor{OutlineG}\ul{Build process intervention}}}

\NewDocumentCommand{\cDeploymentAgent}{o}{\IfValueTF{#1}{\BCircleA{S1}}{\CircleA{S1}}}
\NewDocumentCommand{\cKubernetesOperator}{o}{\IfValueTF{#1}{\BCircleA{S2}}{\CircleA{S2}}}
\NewDocumentCommand{\cSecondaryContainers}{o}{\IfValueTF{#1}{\BCircleA{S3}}{\CircleA{S3}}}
\NewDocumentCommand{\cLifecycleHooks}{o}{\IfValueTF{#1}{\BCircleA{S4}}{\CircleA{S4}}}
\NewDocumentCommand{\cHoneypots}{o}{\IfValueTF{#1}{\BCircleB{HP}}{\CircleB{HP}}}
\NewDocumentCommand{\cReverseProxies}{o}{\IfValueTF{#1}{\BCircleC{RP}}{\CircleC{RP}}}
\NewDocumentCommand{\cFileModifications}{o}{\IfValueTF{#1}{\BCircleD{C1}}{\CircleD{C1}}}
\NewDocumentCommand{\cRuntimeInstrumentation}{o}{\IfValueTF{#1}{\BCircleD{C2}}{\CircleD{C2}}}
\NewDocumentCommand{\cLdPreload}{o}{\IfValueTF{#1}{\BCircleD{C3}}{\CircleD{C3}}}
\NewDocumentCommand{\cPtrace}{o}{\IfValueTF{#1}{\BCircleD{C4}}{\CircleD{C4}}}
\NewDocumentCommand{\cEbpfMonitoring}{o}{\IfValueTF{#1}{\BCircleE{K1}}{\CircleE{K1}}}
\NewDocumentCommand{\cNetfilterRouting}{o}{\IfValueTF{#1}{\BCircleE{K2}}{\CircleE{K2}}}
\NewDocumentCommand{\cEbpfRouting}{o}{\IfValueTF{#1}{\BCircleE{K3}}{\CircleE{K3}}}
\NewDocumentCommand{\cFilesystemImplementations}{o}{\IfValueTF{#1}{\BCircleE{K4}}{\CircleE{K4}}}
\NewDocumentCommand{\cVolumeMounts}{o}{\IfValueTF{#1}{\BCircleF{P1}}{\CircleF{P1}}}
\NewDocumentCommand{\cPolicyEngine}{o}{\IfValueTF{#1}{\BCircleF{P2}}{\CircleF{P2}}}
\NewDocumentCommand{\cCni}{o}{\IfValueTF{#1}{\BCircleF{P3}}{\CircleF{P3}}}
\NewDocumentCommand{\cDeceptionFramework}{o}{\IfValueTF{#1}{\BCircleG{D1}}{\CircleG{D1}}}
\NewDocumentCommand{\cBuildProcess}{o}{\IfValueTF{#1}{\BCircleG{D2}}{\CircleG{D2}}}

\NewDocumentCommand{\mDeploymentAgent}{o}{%
  \hyperref[met:deployment-agent]{\IfValueTF{#1}{\cDeploymentAgent[1]}{\cDeploymentAgent}}}
\NewDocumentCommand{\mKubernetesOperator}{o}{%
  \hyperref[met:kubernetes-operator]{\IfValueTF{#1}{\cKubernetesOperator[1]}{\cKubernetesOperator}}}
\NewDocumentCommand{\mSecondaryContainers}{o}{%
  \hyperref[met:secondary-container]{\IfValueTF{#1}{\cSecondaryContainers[1]}{\cSecondaryContainers}}}
\NewDocumentCommand{\mLifecycleHooks}{o}{%
  \hyperref[met:lifecycle-hooks]{\IfValueTF{#1}{\cLifecycleHooks[1]}{\cLifecycleHooks}}}
\NewDocumentCommand{\mHoneypots}{o}{%
  \hyperref[met:honeypot]{\IfValueTF{#1}{\cHoneypots[1]}{\cHoneypots}}}
\NewDocumentCommand{\mReverseProxies}{o}{%
  \hyperref[met:reverse-proxy]{\IfValueTF{#1}{\cReverseProxies[1]}{\cReverseProxies}}}
\NewDocumentCommand{\mFileModifications}{o}{%
  \hyperref[met:file-modifications]{\IfValueTF{#1}{\cFileModifications[1]}{\cFileModifications}}}
\NewDocumentCommand{\mRuntimeInstrumentation}{o}{%
  \hyperref[met:runtime-instrumentation]{\IfValueTF{#1}{\cRuntimeInstrumentation[1]}{\cRuntimeInstrumentation}}}
\NewDocumentCommand{\mLdPreload}{o}{%
  \hyperref[met:ld-preload]{\IfValueTF{#1}{\cLdPreload[1]}{\cLdPreload}}}
\NewDocumentCommand{\mPtrace}{o}{%
  \hyperref[met:ptrace]{\IfValueTF{#1}{\cPtrace[1]}{\cPtrace}}}
\NewDocumentCommand{\mEbpfMonitoring}{o}{%
  \hyperref[met:ebpf-monitoring]{\IfValueTF{#1}{\cEbpfMonitoring[1]}{\cEbpfMonitoring}}}
\NewDocumentCommand{\mNetfilterRouting}{o}{%
  \hyperref[met:netfilter]{\IfValueTF{#1}{\cNetfilterRouting[1]}{\cNetfilterRouting}}}
\NewDocumentCommand{\mEbpfRouting}{o}{%
  \hyperref[met:ebpf-routing]{\IfValueTF{#1}{\cEbpfRouting[1]}{\cEbpfRouting}}}
\NewDocumentCommand{\mFilesystemImplementations}{o}{%
  \hyperref[met:filesystem-implementations]{\IfValueTF{#1}{\cFilesystemImplementations[1]}{\cFilesystemImplementations}}}
\NewDocumentCommand{\mVolumeMounts}{o}{%
  \hyperref[met:volume-mounts]{\IfValueTF{#1}{\cVolumeMounts[1]}{\cVolumeMounts}}}
\NewDocumentCommand{\mPolicyEngine}{o}{%
  \hyperref[met:policy-management]{\IfValueTF{#1}{\cPolicyEngine[1]}{\cPolicyEngine}}}
\NewDocumentCommand{\mCni}{o}{%
  \hyperref[met:cni]{\IfValueTF{#1}{\cCni[1]}{\cCni}}}
\NewDocumentCommand{\mDeceptionFramework}{o}{%
  \hyperref[met:deception-framework]{\IfValueTF{#1}{\cDeceptionFramework[1]}{\cDeceptionFramework}}}
\NewDocumentCommand{\mBuildProcess}{o}{%
  \hyperref[met:build-process]{\IfValueTF{#1}{\cBuildProcess[1]}{\cBuildProcess}}}

\newcommand{\cSupportingMethods}{\CircleA{A-D}}
\newcommand{\cContainersPods}{\CircleD{G-J}}
\newcommand{\cKernelNetwork}{\CircleE{K-M}}
\newcommand{\cContainerPlatform}{\CircleF{N-P}}
\newcommand{\cDevInteraction}{\CircleG{Q-R}}

\newcommand{\mSupportingMethods}{\hyperref[sec:supporting-methods]{\cSupportingMethods}}
\newcommand{\mContainersPods}{\hyperref[sec:container-pod-methods]{\cContainersPods}}
\newcommand{\mKernelNetwork}{\hyperref[sec:kernel-network-methods]{\cKernelNetwork}}
\newcommand{\mContainerPlatform}{\hyperref[sec:kubernetes-methods]{\cContainerPlatform}}
\newcommand{\mDevInteraction}{\hyperref[sec:developer-interaction-methods]{\cDevInteraction}}

%% file: includes/marks.tex
\newcommand{\DownMark}{$\blacktriangledown$}
\newcommand{\UpMark}{$\blacktriangle$}

\newcommand{\EmptyMark}{{\color{black!10}$\mdlgwhtcircle$}} 
\newcommand{\BadMark}{{\color{black!10}{$\mdlgwhtcircle$}}} 
\newcommand{\PoorMark}{{\color{black!30}{$\circleurquadblack$}}} 
\newcommand{\HalfMark}{{\color{black!45}{$\circlerighthalfblack$}}} 
\newcommand{\FairMark}{{\color{black!45}{$\circlerighthalfblack$}}} 
\newcommand{\GoodMark}{{\color{black!65}{$\blackcircleulquadwhite$}}} 
\newcommand{\FullMark}{$\mdlgblkcircle$} 
\newcommand{\ExcellentMark}{{\color{black!100}{\FullMark}}} 

\newcommand{\YesMark}{\ding{52}}
\newcommand{\NoMark}{{\color{gray!30}\ding{56}}}

\newcommand{\NothingMark}{---} 
\newcommand{\NanMark}{\NothingMark}

\newcommand{\DeplBuiltIn}{{\scriptsize\faCode}}
\newcommand{\DeplAddedTo}{{\scriptsize\faPlusCircle}}
\newcommand{\DeplInFrontOf}{{\scriptsize\faLayerGroup}}
\newcommand{\DeplStandAlone}{{\scriptsize\faHouseUser}}

\newcommand{\LayerApplication}{{\scriptsize\textsc{App}}}
\newcommand{\LayerSystem}{{\scriptsize\textsc{Sys}}}
\newcommand{\LayerNetwork}{{\scriptsize\textsc{Net}}}
\newcommand{\LayerData}{{\scriptsize\textsc{Dat}}}

\newcommand{\DimSource}{{\scriptsize\textsc{Src}}}
\newcommand{\DimPlugins}{{\scriptsize\textsc{Plu}}}
\newcommand{\DimRuntime}{{\scriptsize\textsc{Run}}}
\newcommand{\DimLibraries}{{\scriptsize\textsc{Lib}}}
\newcommand{\DimContainer}{{\scriptsize\textsc{Ctr}}}
\newcommand{\DimKernel}{{\scriptsize\textsc{Ker}}}

\newcommand{\ResKubernetes}{{\scriptsize\textsc{K8s}}}
\newcommand{\ResOciCompliant}{{\scriptsize\textsc{OCI}}}
\newcommand{\ResServiceMesh}{{\scriptsize\textsc{SM}}}
\newcommand{\ResLibc}{{\scriptsize\textsc{libc}}}
\newcommand{\ResManagedRuntime}{{\scriptsize\textsc{MR}}}
\newcommand{\ResNetfilter}{{\scriptsize\textsc{NF}}}
\newcommand{\ResEbpf}{{\scriptsize\textsc{eBPF}}}
\newcommand{\ResPolicyEngine}{{\scriptsize\textsc{PM}}}
\newcommand{\ResCniPlugin}{{\scriptsize\textsc{CNI}}}

%% file: includes/cites.tex
\newcommand{\citecloudbasedhoneypots}{\cite{%
    Memari2014:VirtualHoneynetBased, 
    Kedrowitsch2017:FirstLookUsing, 
    Osman2019:SandnetHighQuality, 
    Gupta2021:HoneyKubeDesigningHoneypot, 
    Gupta2023:HoneyKubeDesigningDeploying, 
    Machmeier2023:HoneypotImplementationCloud, 
    Spahn2023:ContainerOrchestrationHoneypot, 
    Reti2021:EscapeFakeIntroducing, 
    Li2022:OptimalDefensiveDeception, 
    Priya2023:ContainerizedCloudbasedHoneypot, 
    Zambianco2023:ResourceawareCyberDeception 
  }}

\newcommand{\citereverseproxysolutions}{\cite{%
    Han2017:EvaluationDeceptionBasedWeb,
    Araujo2014:PatchesHoneyPatchesLightweight,
    Barron2021:ClickThisNot,
    Fraunholz2018:CloxyContextawareDeceptionasaService,
    Sahin2020:LessonsLearnedSunDEW,
    Pohl2015:HiveZeroConfiguration
  }}

\newcommand{\citesidecaroverheadinresearch}{\cite{%
    Zhu2023:DissectingOverheadsService,
    Elkhatib2023:EvaluationServiceMesh,
    Ganguli2021:ChallengesOpportunitiesPerformance,
    Schober2023:IstioServiceMesh,
    Schneider2023:PerformanceBenchmarkingOpensource
  }}

\newcommand{\citesidecaroverheadinindustry}{\cite{%
    Graf2021:HowEBPFWill,
    Howard2022:IntroducingAmbientMesh
  }}

\newcommand{\citemalwaredeceptionframeworks}{\cite{%
    Islam2021:CHIMERAAutonomousPlanning,
    Sajid2021:SODASystemCyber,
    Sajid2020:DodgeTronAutonomousCyber,
    Alsaleh2018:GExtractorAutomatedExtraction
  }}

\newcommand{\citehoneypotfingerprinting}{\cite{%
    Srinivasa2023:GottaCatchEm,
    Mukkamala2007:DetectionVirtualEnvironments
  }}

%% file: content/01-introduction.tex
\section{Introduction}
\label{sec:introduction}

There is no shortage of research on the benefits of cyber deception:%
~\cite{Ferguson-Walter2021:ExaminingEfficacyDecoybased,Ferguson-Walter2023:CyberExpertFeedback}
Techniques such as honeypots or honeytokens are effective at
slowing-down and deterring adversaries,
providing threat intelligence,
and improving incident detection and response.
%
What remains a barrier to widespread adoption of cyber deception technology
is its deployment in real-world software systems.
Lance Spitzner, one of the first to study honeypots%
~\cite{Spitzner2002:HoneypotsTrackingHackers,Spitzner2003:HoneypotsCatchingInsider}
and honeytokens~\cite{Spitzner2003:HoneytokensOtherHoneypot},
commented in \citeyear{Chuvakin2019:WillDeceptionFizzle},
three decades after their inception,
that cyber deception was held back not by the concept, but by technology:
``Every honeypot back then had to be crafted, customized and managed by hand.''%
~\cite{Chuvakin2019:WillDeceptionFizzle}.
Recent advances in virtualization technology have improved the applicability
of ``classic'' honeypots~\cite{Zhang2021:ThreeDecadesDeception}, but
the practical application of deception techniques that are closely intertwined with applications
(e.g., fully automated honeytokens in file systems) has not yet been fully realized.
Intertwining deception techniques tightly with software systems
shall increase their effectiveness compared to self-contained honeypots%
~\cite{Han2017:EvaluationDeceptionBasedWeb},
which are easier to discern from genuine assets%
~\citehoneypotfingerprinting{}.
Modern research on cyber deception and moving target defense
mimics characteristics of real data~\cite{Bercovitch2011:HoneyGenAutomatedHoneytokens}
and strives towards dynamic and personalized deception%
~\cite{%
  Gonzalez2020:DesignDynamicPersonalized,
  Niakanlahiji2020:HoneyBugPersonalizedCyber,
  Dowling2020:NewFrameworkAdaptive}.
Such new concepts often require the insertion of decoys
into already built software applications, as well as regular customization.
These requirements are most critical when deception technology is provided ``as a service``%
~\cite{%
  Fraunholz2018:CloxyContextawareDeceptionasaService,
  Araujo2016:EngineeringCyberDeceptiveSoftware}.


\citeauthor{Han2018:DeceptionTechniquesComputer}
classify the layer of deception into network, system, application and data%
~\cite{Han2018:DeceptionTechniquesComputer}.
Our work focuses on the application layer, that is, techniques that are
``linked to specific classes of applications, such as web applications''.
This perspective also harmonizes well with modern service-oriented architectures%
~\cite{Gotz2018:ChallengesProductionMicroservices}
and cloud-native paradigms~\cite{CloudNativeComputingFoundation2024:CloudNativeDefinition}.
%
Container orchestration platforms such as Kubernetes
are prevalent not only in modern software systems,
but also in recent research on cloud-based honeypots~\citecloudbasedhoneypots{}.
Therefore, our work includes methods that are specifically suited for Kubernetes.
Four methods are exclusive
to Kubernetes; most are applicable to any computer system.%
\footnote{
  These methods (\S\ref{sec:methods}) are exclusive to Kubernetes:
  \mKubernetesOperator{}~\tKubernetesOperator{},
  \mSecondaryContainers{}~\tSecondaryContainers{},
  \mPolicyEngine{}~\tPolicyEngine{},
  and \mCni{}~\tCni{}.
}
%
%
Our contributions are:
\begin{enumerate}
  \item A technical overview of \numMethods{}~methods to achieve application layer cyber deception,
        given already built software artifacts without source code.
  \item A qualitative evaluation of these on
        technical, topological, operational, and efficacy properties.
\end{enumerate}

%% file: content/02-problem.tex
\section{Problem Scope}
\label{sec:problem-statement}

We demonstrate three typical use cases for application layer cyber deception%
~\cite{Kahlhofer2020:ReconstructingMultiStepCyber},
given only built artifacts of an application (no source code access),
in the context of cloud-native environments (represented by Kubernetes).


\textbf{Application layer cyber deception.}
(A)~\emph{Decoy request.} We want to add the code in Listing~\ref{lst:network-decoy}
to the HTML payload of an HTTP response.
If adversaries probe the associated application for vulnerabilities,
we expect them to waste time in exploiting this endpoint,
because it seems to carry a path traversal weakness. 
We then need to detect access attempts to this endpoint.
(B)~\emph{Fake API.} To make the previous use case more interactive,
we want to patch an ``/admin/api'' endpoint into our application
-- if it does not already exist -- and respond to it with some deceptive payload.
(C)~\emph{Honeytoken.} Suppose an adversary has managed to gain access
to a (container) file system. We want to place files that appear sensitive
(e.g., a ``service-token'') and detect access attempts to them.

\begin{lstlisting}[
  label=lst:network-decoy,
  caption={JavaScript code to issue a suspicious HTTP request.}
]
const req = new XMLHttpRequest();
req.open("GET", "/admin/api?p=../config.json");
req.send();
\end{lstlisting}

\textbf{No developer interaction.}
Software applications are rarely deployed by the team that wrote the code,
and often the responsibility for security measures lies entirely elsewhere.
While this situation is improving
~\cite{Ebert2016:DevOps,Beetz2022:GitOpsEvolutionDevOps},
it is valuable to explore how we can intertwine defensive cyber deception 
with applications without having control over the entire software development lifecycle.
The desire to add cyber deception on top of an existing system becomes increasingly strong in
large organizations, where security operators want to add cyber deception
-- among other security measures -- consistently across hundreds of applications,
following a unified process.
We assume that we only have access to built artifacts,
which are typically distributed as container images.
This adds a few technical challenges that are rarely addressed in the literature,
e.g., dealing with a wide variety
of software technologies
~\cite{Sahin2020:LessonsLearnedSunDEW},
limited access to source code, and compatibility issues
~\cite{Araujo2020:ImprovingCybersecurityHygiene}.

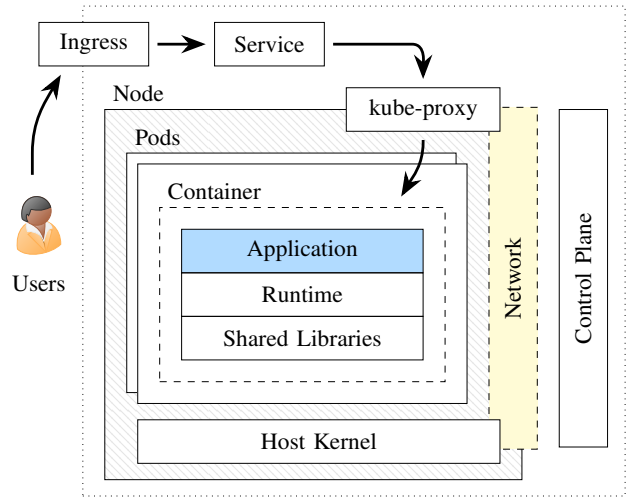
\begin{figure}[t]
  \centering
  \resizebox{\columnwidth}{!}{\input{./figures/environment.tex}}
  \caption{Components of a typical Kubernetes cluster.}
  \label{fig:environment}
\end{figure}

\textbf{Cloud-native platforms.}
Cloud-native paradigms, that is, building loosely coupled systems combined with robust automation%
~\cite{CloudNativeComputingFoundation2024:CloudNativeDefinition},
harmonize well with the requirements of modern, dynamic, adaptive cyber deception techniques.
Thus, we include methods that are applicable to Kubernetes, 
which is the dominant platform for modern software development. 
Figure~\ref{fig:environment} shows its typical components:
A \emph{pod} represents the actual workload (the software application).
A pod consists of one or more containers that share storage and network resources.
Their \emph{container images} contain the built software artifacts.
A \emph{node} is the physical or virtual machine that runs these pods.
A \emph{cluster} is formed from one or more nodes.
Unique to Kubernetes is the so-called \emph{control plane},
which administers the cluster. Parts of it can also be customized, e.g.,
the \emph{admission control} can be set to modify the \emph{manifest} of new workloads
(e.g., mounted volumes, environment variables) before workloads are added to the cluster
(see \mKubernetesOperator{}~\tKubernetesOperator{}).
Workloads are often addressed indirectly through \emph{services},
which provide a network-accessible endpoint for some business logic.
%
An \emph{ingress} is a special kind of pod at the perimeter
that provides connectivity to the outside world.
The underlying workload of ingresses are often reverse proxies, such as
NGINX~\cite{Sysoev:NGINX} or Envoy\cite{TheEnvoyProjectAuthors:Envoy}.

\FloatBarrier

%% file: figures/environment.tex
{\scalefont{1.0}\begin{tikzpicture}[scale=0.66,on grid]
  \def\podoffset{.25}%
  \def\arrowoffset{.2}%
  \def\yoffset{1}%
  \def\boxheight{1}%
  \def\boxwidth{1.1}%

  \coordinate (ClusterTL) at (-6,4.25+.125);
  \coordinate (ClusterBR) at (6.35,-6.75-.125); 
  \coordinate (NodeTL) at (-5.5,2);
  \coordinate (NodeBR) at (4,-6.5); 

  \coordinate (KProxyTR) at (3.25+\podoffset,2+0.5*\boxheight);
  \coordinate (KProxyBL) at ($(KProxyTR)+(-3.5,-\boxheight)$);

  \coordinate (IngressTL) at (-7,4);
  \coordinate (IngressBR) at ($(IngressTL)+(2.5,-\boxheight)$);

  \coordinate (StickT) at (-7, 0);

  \coordinate (ServiceTL) at (-3,4);
  \coordinate (ServiceBR) at ($(ServiceTL)+(2.5,-\boxheight)$);

  \coordinate (PodTL) at (-5+\podoffset,2-\yoffset-\podoffset);
  \coordinate (PodBR) at (2.5+\podoffset,-4.75);
  \coordinate (PodBgTL) at ($(PodTL)+(-\podoffset,\podoffset)$);
  \coordinate (PodBgBR) at ($(PodBR)+(-\podoffset,\podoffset)$);

  \coordinate (CtrTL) at (-4.5+\podoffset,2-2*\yoffset-\podoffset);
  \coordinate (CtrBR) at (2+\podoffset,2-3*\yoffset-\podoffset-3*\boxheight);

  \coordinate (AppTL) at ($(CtrTL)+(0.5,-0.5)$);
  \coordinate (AppBR) at ($(AppTL)+(5.5,-\boxheight)$);
  \coordinate (RuntimeTL) at ($(AppTL)+(0,-\boxheight)$);
  \coordinate (RuntimeBR) at ($(RuntimeTL)+(5.5,-\boxheight)$);
  \coordinate (LibsTL) at ($(RuntimeTL)+(0,-\boxheight)$);
  \coordinate (LibsBR) at ($(LibsTL)+(5.5,-\boxheight)$);

  \coordinate (KernelTL) at (-5+\podoffset,-5.25+.125);
  \coordinate (KernelBR) at (3.25+\podoffset,-5.25+.125-\boxheight);

  \coordinate (CtrlTL) at (4.85,2);
  \coordinate (CtrlBR) at (4.85+\boxwidth,-5.25-0.5*\boxheight);

  \coordinate (NetworkTL) at (3.25,2+0.05); 
  \coordinate (NetworkBR) at (3.25+\boxwidth,-5.25-0.5*\boxheight-0.05);

  \draw[dotted] (ClusterTL) rectangle (ClusterBR) node[midway,align=center]{};

  \draw[pattern=north west lines,pattern color=black!10]
  (NodeTL) rectangle (NodeBR) node[midway,align=center]{};
  \node[above right] at (NodeTL) {Node};

  \draw[fill=white] (PodBgTL) rectangle (PodBgBR) node[midway,align=center]{};
  \draw[fill=white] (PodTL)[fill=white] rectangle (PodBR) node[midway,align=center]{};
  \node[above right] at (PodBgTL) {Pods};

  \draw[dashed] (CtrTL) rectangle (CtrBR) node[midway,align=center]{};
  \node[above right] at (CtrTL) {Container};

  \draw[fill=SteelBlue1!50] (AppTL) rectangle (AppBR) node[midway,align=center]{Application};
  \draw (RuntimeTL) rectangle (RuntimeBR) node[midway,align=center]{Runtime};
  \draw (LibsTL) rectangle (LibsBR) node[midway,align=center]{Shared Libraries};

  \draw (CtrlTL) rectangle (CtrlBR)
  node[midway,align=center,rotate=90] {Control Plane};
  \draw[fill=yellow!20,dashed] (NetworkTL) rectangle (NetworkBR)
  node[midway,align=center,rotate=90] {Network};

  \draw[fill=white] (KernelTL) rectangle (KernelBR) node[midway,align=center]{Host Kernel};

  \draw[fill=white] (KProxyTR) rectangle (KProxyBL) node[midway,align=center]{kube-proxy};
  \draw[fill=white] (IngressTL) rectangle (IngressBR) node[midway,align=center]{Ingress};
  \draw[fill=white] (ServiceTL) rectangle (ServiceBR) node[midway,align=center]{Service};

  \draw[-{Stealth[scale=1.25]},very thick] ($(IngressBR)+(\arrowoffset,0.5*\boxheight)$)
  -- ($(ServiceTL)+(-\arrowoffset,-0.5*\boxheight)$);
  \draw[-{Stealth[scale=1.25]},very thick] ($(ServiceBR)+(\arrowoffset,0.5*\boxheight)$)
  -- ($(ServiceBR)+(2,0.5*\boxheight)$) to[out=0,in=90] ($(KProxyTR)+(-.5*3.5,\arrowoffset)$);
  \draw[-{Stealth[scale=1.25]},very thick] ($(KProxyBL)+(0.5*3.5,-\arrowoffset)$)
  -- ($(KProxyBL)+(0.5*3.5,-0.25)$) to[out=-90,in=60] ($(CtrTL)+(5.5,.25)$);
  \draw[-{Stealth[scale=1.25]},very thick] ($(StickT)+(-.15,.5)$)
  -- ($(StickT)+(-.15,1)$) to[out=90,in=240] ($(IngressTL)+(.5,-\boxheight-\arrowoffset)$);

  \node[alice,anchor=north,scale=1.75] at (StickT){};
  \node[anchor=center] at ($(StickT)+(0,-2)$){Users};
\end{tikzpicture}}

%% file: content/03-literature.tex
\section{Related Work, Surveys, and Taxonomies}
\label{sec:related-work}

The concept of honeypots%
~\cite{Spitzner2003:HoneypotsCatchingInsider,Spitzner2003:HoneytokensOtherHoneypot}
and honeytokens~\cite{Spitzner2003:HoneytokensOtherHoneypot} was originally described
by \citeauthor{Spitzner2003:HoneypotsCatchingInsider} as ``an information system resource
whose value lies in unauthorized or illicit use of that resource''%
~\cite{Spitzner2003:HoneypotsCatchingInsider}.

Many surveys, taxonomies, and classifications on cyber deception have been introduced%
~\cite{%
  Han2018:DeceptionTechniquesComputer,
  Zhang2021:ThreeDecadesDeception,
  Fraunholz2018:DemystifyingDeceptionTechnology,
  Fan2018:EnablingAnatomicView,
  Javadpour2024:ComprehensiveSurveyCyber,
  Efendi2019:SurveyDeceptionTechniques,
  Lu2020:CyberDeceptionComputer,
  Mohan2022:LeveragingComputationalIntelligence,
  Qin2023:HybridCyberDefense,
  Urias2017:TechnologiesEnableCyber,
  Almeshekah2014:PlanningIntegratingDeception,
  Bringer2012:SurveyRecentAdvances,
  Mokube2007:HoneypotsConceptsApproaches,
  Rowe2004:TwoTaxonomiesDeception,
  Scottberg2002:InternetHoneypotsProtection,
  Whaley1982:GeneralTheoryDeception},
but few authors focused on technical aspects
of \mHoneypots{}~\tHoneypots{} and their deployment%
~\cite{Nawrocki2016:SurveyHoneypotSoftware,Fraunholz2017:DeploymentStrategiesDeception} and yet
none focused exclusively on the application layer and methods that require no developer interaction.
\citeauthor{Han2018:DeceptionTechniquesComputer} also noted that
``[most works] focused on the introduction of new deceptive techniques and elements,
but [...] rarely discussed where \emph{(and how)} such elements should be placed in the system''%
~\cite{Han2018:DeceptionTechniquesComputer}.
The inventors of honeypatches%
~\cite{%
  Araujo2014:PatchesHoneyPatchesLightweight,
  Araujo2015:ExperiencesHoneyPatchingActive,
  Araujo2016:EmbeddedHoneypotting},
which are silently-patched application layer vulnerabilities
that still seem exploitable on the surface,
also noted that ``the concept of honey-patching is straightforward,
realizing it in practice is not''~\cite{Araujo2014:PatchesHoneyPatchesLightweight}.

Cyber deception is often studied using game theory
~\cite{Zhu2021:SurveyDefensiveDeception,Pawlick2019:GametheoreticTaxonomySurvey}.
Methods that dynamically adapt deception techniques to an adversary's behavior 
have been well-studied from a theoretical standpoint%
~\cite{Rass2020:CyberSecurityCriticalInfrastructures}.
However, these models mostly assume a set of known deception strategies that are randomly
interchangeable, but do not necessarily specify how to technically realize a game move.
%

\mRuntimeInstrumentation{}~\tsRuntimeInstrumentation{},
the \mLdPreload{}~\tLdPreload{} trick, and tracing with \mPtrace{}~\tPtrace{}
are technical methods to hook and patch functions, which is well studied
~\cite{Lopez2017:SurveyFunctionSystem,Islam2023:RuntimeSoftwarePatching}.
Hooking-based approaches are commonly found in the domain of malware analysis,
where malware is deceived with intercepted API methods
~\citemalwaredeceptionframeworks{}.
Custom \mFilesystemImplementations~\tFilesystemImplementations{}
are also used to detect ransomware specifically%
~\cite{%
  Kharaz2016:UNVEILLargeScaleAutomated,
  Continella2016:ShieldFSSelfhealingRansomwareaware,
  Sheen2022:RSentryDeceptionBased
}.


Most work on application layer cyber deception
places \mReverseProxies{}~\tReverseProxies{} in-front of applications%
~\citereverseproxysolutions{}.
%
\citeauthor{Reti2023:HoneyInfiltratorInjecting}~\cite{Reti2023:HoneyInfiltratorInjecting}
used \mNetfilterRouting{}~\tNetfilterRouting{} in combination with Scapy%
~\cite{Biondi:Scapy},
which is a user-space process for packet manipulation.
\citeauthor{Kern2024:InjectingSharedLibraries}~\cite{Kern2024:InjectingSharedLibraries}
used \mLdPreload{}~\tLdPreload{} to modify HTTP packet headers for deception.
\citeauthor{Araujo2020:ImprovingCybersecurityHygiene}%
~\cite{Araujo2020:ImprovingCybersecurityHygiene}
patched applications by injecting (and just-in-time compiling)
code into running processes with \mPtrace{}~\tPtrace{}.
Deploying honeypots as containers and in cloud platforms has been well studied%
~\citecloudbasedhoneypots{}.
Notably, \citeauthor{Gupta2023:HoneyKubeDesigningDeploying}%
~\cite{Gupta2023:HoneyKubeDesigningDeploying,Gupta2021:HoneyKubeDesigningHoneypot}
introduced ``HoneyKube'' to orchestrate deployment and monitoring
of self-contained \mHoneypots{}~\tHoneypots{} in Kubernetes.

Research on application layer cyber deception also focuses on quantifying
the enticingness of novel techniques and automating the generation of enticing payloads%
~\cite{%
  Sahin2022:ApproachGenerateRealistic,
  Sahin2022:MeasuringDevelopersWeb,
  Sahin2020:LessonsLearnedSunDEW,
  Bercovitch2011:HoneyGenAutomatedHoneytokens,
  Bowen2010:AutomatingInjectionBelievable
}.
The effectiveness of cyber deception and its
psychological aspects are also extensively studied%
~\cite{%
  Ferguson-Walter2023:CyberExpertFeedback,
  Ferguson-Walter2019:TularosaStudyExperimental,
  Ferguson-Walter2021:ExaminingEfficacyDecoybased,
  Ferguson-Walter2020:EmpiricalAssessmentEffectiveness,
  Ferguson-Walter2019:WorldCTFNot,
  Huang2022:ADVERTAdaptiveDataDriven,
  Cranford2018:LearningCyberDeception,
  Cranford2021:CognitiveTheoryCyber,
  Cranford2020:AdaptiveCyberDeception,
  Gutzwiller2018:OhLookButterfly,
  Gabrys2023:EmotionalStateClassification,
  Gonzalez2020:DesignDynamicPersonalized,
  BenSalem2011:DecoyDocumentDeployment}.


%
%
%
%

%% file: content/04-properties.tex
\section{Properties of Technical Methods}
\label{sec:properties}

\begin{figure*}[t]
  \centering
  {\small\input{./figures/properties.tex}}
  \caption{Properties of technical methods for application layer cyber deception.}
  \label{fig:properties}
\end{figure*}
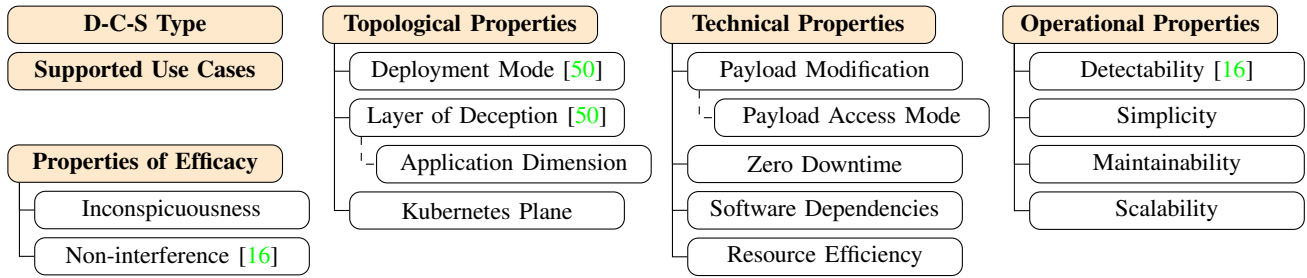

We categorize methods by their type, properties (technical, topological, operational),
efficacy, and suitability to achieve our three representative use cases
(Figure~\ref{fig:properties}). We map properties to methods in Table~\ref{tab:overview}.
Reasons for excluding some properties are found in
Appendix~\ref{sec:appendix:excluded-properties}.

\textbf{D-C-S Type.} 
\citeauthor{Fan2018:EnablingAnatomicView}~\cite{Fan2018:EnablingAnatomicView}
proposed to differentiate \emph{decoys} and \emph{captors} (D-C).
Decoys are the entities that are attacked (e.g., honeypots),
while captors perform security-related functions (e.g., logging attacks).
We adopt this distinction and further differentiate \emph{supporting} methods,
which assist in deploying or configuring a decoy or captor,
but are not themselves decoys or captors.

\phantomsection
\label{par:topological-properties}
\textbf{Topological Properties.}
\citeauthor{Han2018:DeceptionTechniquesComputer}~\cite{Han2018:DeceptionTechniquesComputer}
classify deception techniques into four dimensions:
their \emph{goal} (e.g., detection of attacks),
their \emph{unit} (e.g., service, vulnerability)
-- as originally introduced by \citeauthor{Almeshekah2014:PlanningIntegratingDeception}%
~\cite{Almeshekah2014:PlanningIntegratingDeception} --,
their \emph{deployment mode},
and the \emph{layer} of deception.
We adopt the deployment mode and layer dimensions,
but not the goal dimension, which we see as independent of technical methods,
nor the unit dimension, which we see as implementation-specific.

They introduced the following \emph{deployment modes}:%
~\cite{Han2018:DeceptionTechniquesComputer}
\begin{itemize}
  \item \DeplBuiltIn~\emph{Built-in}
        means to add something at the design phase of an application,
        e.g., in source code, which typically requires developer interaction.
  \item \DeplInFrontOf~\emph{In-front of}
        means to interpose something between an adversary and a genuine application,
        e.g., with proxies, filters, or (function) hooks.
  \item \DeplAddedTo~\emph{Added-to}
        means to include, integrate, or modify something at runtime,
        e.g., adding files to file systems, but also pods or configuration to a cluster.
  \item \DeplStandAlone~\emph{Stand-alone}
        methods are separated and isolated from target (production) systems.
\end{itemize}

They introduced the following \emph{layers}:~\cite{Han2018:DeceptionTechniquesComputer}
\begin{itemize}
  \item \emph{Network layer} methods (e.g., typical honeypots) are
        ``accessible over the network and [...]
        are not bound to any specific host configuration''%
        ~\cite{Han2018:DeceptionTechniquesComputer}
  \item \emph{System layer} methods are bound to hosts. We consider Kubernetes nodes
        and the entire platform to be part of this layer, so this layer also includes
        kernel-based techniques and cluster-wide policies.
  \item \emph{Application layer} methods are bound to specific applications.
        We consider containers, pods and their file system to be part of this layer.
  \item The \emph{data layer} leverages data, documents, or files.
\end{itemize}

We further subdivide the application layer into six finer levels,
inspired by the ones introduced by 
\citeauthor{Islam2023:RuntimeSoftwarePatching}~\cite{Islam2023:RuntimeSoftwarePatching}:
\begin{itemize}
  \item \emph{Source code level.}
        Methods that directly modify the source code, probably at the design phase.
  \item \emph{Plugin system level.}
        Applications that provide means for third parties to easily add custom code.
  \item \emph{Runtime level.}
        Reflection and introspection capabilities provided by managed languages
        such as the JVM Tool Interface (JVM TI)~\cite{Oracle2013:JVMTMTool}.
  \item \emph{Library level.}
        Shared libraries, which are typically linked on application startup.
  \item \emph{Container level.}
        Methods that operate from within a typical container namespace, seeing resources
        such as \emph{processes} and their (network) interfaces (e.g., listening HTTP sockets)
        and the \emph{file system} (and application artifacts or auxiliary files).
  \item \emph{Kernel level.}
        Operating system features, e.g. kernel modules,
        eBPF programs\cite{Rice2022:WhatEBPF},
        or device drivers.
\end{itemize}
We also differentiate where a method is used in a cloud-native platform.
This is important for system operators because
they are often only authorized to maintain some parts of a cluster.
Note also that the location where a method is employed (e.g., a central configuration)
may be different from the location it affects (e.g., a workload on a node).
We differentiate:
\begin{itemize}
  \item \emph{Data plane} components, i.e., methods employed at cluster nodes,
        affecting pods or containers.
        %
  \item \emph{Control plane} components, e.g., methods that touch the central configuration,
        admission control, the scheduler, the API server, or operators.
\end{itemize}

\textbf{Technical Properties.}
We find the following technical properties relevant to our problem statement.
The first two can be seen as sub-dimensions
of \emph{adaptability}~\cite{Fan2018:EnablingAnatomicView}:
\begin{itemize}
  \item \emph{Payload modification},
        i.e., the ability to \emph{modify or redirect application data}.
        Deception methods beyond self-contained honeypots 
        need to modify response payloads, install new endpoints, or modify file contents.
        This can be done by \emph{directly} modifying code or data,
        or \emph{indirectly}, e.g., with a proxy. 
        We typically prefer indirect modification, as we want to be 
        ``invisible'' to adversaries~\cite{Reti2023:HoneyInfiltratorInjecting}.
  \item \emph{Zero downtime}, i.e., 
        if a method requires a restart of an application when first installed
        (potentially resulting in a short service downtime), and also, if a restart is required
        to reconfigure or adapt an already deployed deception method.
        The latter is a critical requirement to realize dynamic and adaptive deception%
        ~\cite{Jajodia2011:MovingTargetDefense,Rass2017:CostGamePlaying},
        but also a strategy in game-theoretic defense models against intrusions%
        ~\cite{Rass2023:GametheoreticAPTDefense,Lei2017:OptimalStrategySelection}.
  \item Few \emph{software dependencies}
        and compatibility with many programming languages and runtimes. 
  \item \emph{Resource efficiency}, i.e.,
        low CPU and memory overhead and fast response times
        compared to an unmodified application.
        Since increased response times are also noticeable to adversaries%
        ~\cite{Mukkamala2007:DetectionVirtualEnvironments},
        it is critical not to reveal a method by that.
\end{itemize}

\textbf{Operational Properties.}
We assume that people responsible for maintaining or relying on a cyber deception method
(e.g., for incident detection and response), care about at least the following properties:
\begin{itemize}
  \item How well the method is able to \emph{detect} attacks and generate alerts.
        \citeauthor{Bowen2009:BaitingAttackersUsing}~\cite{Bowen2009:BaitingAttackersUsing}
        introduced this as a necessary property for all decoys.
  \item The \emph{technical simplicity} of the method.
  \item How much \emph{maintenance} the method requires.
  \item How well the method \emph{scales} to large systems.
\end{itemize}

\textbf{Properties of Efficacy.}
The efficacy of a complete deception strategy may depend on the following:
\begin{itemize}
  \item \emph{Inconspicuousness}, i.e.,
        how obvious or clearly visible a deception method is to an adversary.
        We treat \emph{fingerprintability}%
        ~\cite{Gupta2021:HoneyKubeDesigningHoneypot,Gupta2023:HoneyKubeDesigningDeploying}, i.e.,
        attacker's ability to discern deceptive from genuine assets,
        as a sub-dimension of that.
        Inconspicuousness is not to be confused with the 
        \emph{conspicuousness of decoys}~\cite{Bowen2009:BaitingAttackersUsing},
        which instead describes how obvious a concrete deception element is, not the method.
  \item \emph{Non-interference} with genuine assets of an application%
        ~\cite{Bowen2009:BaitingAttackersUsing,Reti2023:HoneyInfiltratorInjecting},
        and its \emph{differentiability} to legitimate users.
        Since differentiability typically correlates with non-interference%
        ~\cite{BenSalem2011:DecoyDocumentDeployment}, we combine them.
        In our evaluation, we also combine this with a measure of how well methods avoid
        increasing the attack surface of an application, i.e.,
        ``it must not make the system less secure''~\cite{Reti2023:HoneyInfiltratorInjecting}.
        \citeauthor{Gupta2023:HoneyKubeDesigningDeploying}%
        ~\cite{Gupta2021:HoneyKubeDesigningHoneypot,Gupta2023:HoneyKubeDesigningDeploying}
        denoted this aspect as \emph{security measures for decoys}.
\end{itemize}

%% file: figures/properties.tex
\tikzset{%
  parent/.style={align=center,text width=3cm,rounded corners=3pt},
  child/.style={align=center,text width=3cm,rounded corners=3pt}
}

\forestset{head/.style={fill=ACMOrange!20,font=\bfseries}}
\forestset{highlight/.style={fill=ACMBlue!20}}
\forestset{leaf/.style={draw=none,no edge,color=gray,font=\footnotesize}}

\begin{forest}
  for tree={%
  node options={align=center,},
  rounded corners,
  draw,
  minimum height=14pt,
  text width=3.4cm,
  s sep=3pt,
  },
  where level=0{%
  s sep=8pt
  }{%
  folder,
  grow'=0,
  if level=1{%
  before typesetting nodes={child anchor=north},
  edge path'={(!u.parent anchor) -- ++(0,-5pt) -| (.child anchor)},
  }{}
  }
  [,phantom
  [D-C-S Type,head
  [Supported Use Cases,head,xshift=-10pt,no edge]
  [,phantom]
  [Properties of Efficacy,head,xshift=-10pt,no edge
    [Inconspicuousness,xshift=-10pt]
    [Non-interference~\cite{Bowen2009:BaitingAttackersUsing},xshift=-10pt]
  ]
  ]
  [Topological Properties,head
  [Deployment Mode~\cite{Han2018:DeceptionTechniquesComputer}
  ]
  [Layer of Deception~\cite{Han2018:DeceptionTechniquesComputer}
  [Application Dimension,edge={dashed}
  ]
  ]
  [Kubernetes Plane
  ]
  ]
  [Technical Properties,head
  [Payload Modification
    [Payload Access Mode,edge={dashed}
    ]
  ]
  [Zero Downtime
  ]
  [Software Dependencies
  ]
  [Resource Efficiency
  ]
  ]
  [Operational Properties,head
  [Detectability~\cite{Bowen2009:BaitingAttackersUsing}]
  [Simplicity]
  [Maintainability]
  [Scalability]
  ]
  ]
\end{forest}

%% file: content/05-methods.tex
\section{Technical Methods}
\label{sec:methods}

\begin{figure*}[t]
  \centering
  \resizebox{\textwidth}{!}{\input{./figures/overview.tex}}
  \caption{A topological overview of the \numMethods{}~technical methods that we present.}
  \label{fig:methods}
\end{figure*}
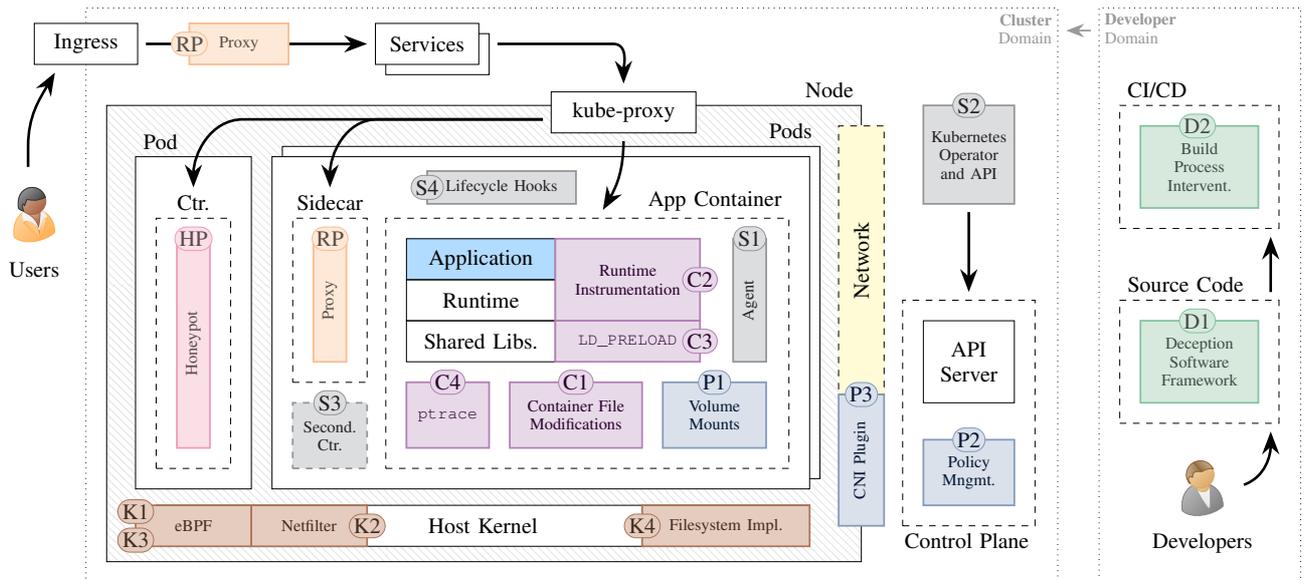

We primarily categorize methods according to
their (virtual) location in a (cloud) platform (Figure~\ref{fig:methods}), 
as this corresponds well with the perspective of system operators:
%
%
\begin{itemize}
  \item \CircleA{S} Supporting methods (\S\ref{sec:supporting-methods})
  \item \CircleB{HP} Self-contained honeypots (\S\ref{sec:isolated-honeypots})
  \item \CircleC{RP} Reverse proxies (\S\ref{sec:reverse-proxies})
  \item \CircleD{C} Containers and pods (\S\ref{sec:container-pod-methods})
  \item \CircleE{K} Kernel and network (\S\ref{sec:kernel-network-methods})
  \item \CircleF{P} Container platform (\S\ref{sec:kubernetes-methods})
  \item \CircleG{D} Methods \emph{with} developer interaction
        (\S\ref{sec:developer-interaction-methods})
\end{itemize}
This categorization shows good consistency, although,
we acknowledge that some supporting methods are also container platform-specific,
and that some kernel-specific methods may also fit container- and pod-specific methods.

\subsection{Supporting Methods}
\label{sec:supporting-methods}


\subsubsection[Deployment Agent]{\cDeploymentAgent~Deployment Agent}
\label{met:deployment-agent}
The most ordinary (supporting) method is to create
and deploy a process (or, often called an agent)
that then places decoys and captors in a system.
The term \emph{agent} is also commonly used in other methods as well
(e.g., \mRuntimeInstrumentation{}~\tRuntimeInstrumentation{} agents),
although they are not \emph{supporting} the deployment or configuration of said methods.
A deployment agent can
freely perform \mFileModifications{}~\tFileModifications{} to create honeytokens,
or use container- and pod-specific methods~(\S\ref{sec:container-pod-methods})
to modify applications and data. With sufficient privileges,
kernel- and network-specific methods~(\S\ref{sec:kernel-network-methods})
are also available to this agent.

Running a process next to an application carries little risk of interference
and provides great flexibility in implementing deception scenarios.
A deployment agent can be installed directly inside the application container
(requiring a \mKubernetesOperator{}~\tKubernetesOperator{},
\mSecondaryContainers{}~\tSecondaryContainers{},
or \mLifecycleHooks{}~\tLifecycleHooks{}
to bring the agent inside the container)
or on the container host (i.e., the node).
A direct deployment within a container keeps the agent close to and isolated from the application,
and is also more robust because Kubernetes then manages the lifecycle of that deployment artifact.
Instead, deploying the agent on the container host would require a broader set of permissions
so that it can still interact with the Linux namespaces of individual containers.
The latter would keep the agent hidden from an adversary that compromised the container,
since otherwise an adversary could easily detect its presence by listing all running processes.

\subsubsection[Kubernetes Operator and API]{\cKubernetesOperator~Kubernetes Operator and API}
\label{met:kubernetes-operator}

Controller and Operator patterns~\cite{Ibryam2019:KubernetesPatterns}
are well-defined ways to add custom, decoupled resources to a Kubernetes cluster
that can also hook into controlling processes.
This is similar to having a \mDeploymentAgent{}~\tDeploymentAgent{},
but it is deployed in its own pod and has pre-set access to the Kubernetes API.

To place a honeytoken in a container's file system,
an operator first configures a \textit{mutating admission webhook}
through the Kubernetes API.
Before new resources are deployed to the cluster,
the operator's webhook is invoked and can mutate the resource definition.
Honeytokens can be added to containers by configuring
\mVolumeMounts{}~\tVolumeMounts{} for them that contain honeytokens.

For already running containers, the operator can also
execute commands in containers via the Kubernetes API~\cite{Erol2019:HowDoesKubectl},
which allows this method to directly add files at runtime,
install a \mDeploymentAgent{}~\tDeploymentAgent{} in a container without restarting it,
or -- assuming sufficient privileges -- execute
container- and pod-specific methods~(\S\ref{sec:container-pod-methods})
or even kernel- and network-specific methods~(\S\ref{sec:kernel-network-methods})
directly.

\subsubsection[Secondary Containers]{\cSecondaryContainers~Secondary Containers}
\label{met:secondary-container}

Kubernetes can add special kinds of containers next to application containers.

\emph{Init containers},
which are meant to run setup scripts,
will always run before the application containers,
and the application containers will not start before their completion.
They use their own container image and cannot access files from the application image,
but can share the same \mVolumeMounts{}~\tVolumeMounts{} with the application container.
An init container can place honeytokens in a volume that will later
be mounted by the application container.
Init containers can be defined
via a \mKubernetesOperator{}~\tKubernetesOperator{}
or \mPolicyEngine{}~\tPolicyEngine{},
but do not take effect until the next time the pod is restarted.

\emph{Ephemeral containers},
which are meant for debugging running applications,
provide a way to dynamically launch a container in pods that are already running.
They can then interact with application processes and their file system
and implement deception methods at runtime.

A \emph{sidecar container} is a similar concept in this category,
which can be used to log events or monitor the main application.
We typically use them to place \mReverseProxies{}~\tReverseProxies{}
in-front of application containers.

\subsubsection[Container Lifecycle Hooks]{\cLifecycleHooks~Container Lifecycle Hooks}
\label{met:lifecycle-hooks}

Most container orchestration platforms provide the ability to run
scripts on container startup (pre-start and post-start hooks) and shutdown (post-stop hook).
This is not only available in Kubernetes,
but is also standardized in the OCI Runtime Specification%
~\cite{OpenContainersInitiative2023:OpenContainerInitiative}.
These platforms also typically allow to override the \emph{entrypoint} of a container,
i.e., the command or process that is run at start.

Such hooks can perform deception methods similar
to what a \mDeploymentAgent{}~\tDeploymentAgent{} can do. 
As with~\mSecondaryContainers{}~\tSecondaryContainers{},
such hooks must first be set in the manifest with
a \mKubernetesOperator{}~\tKubernetesOperator{}
or by a \mPolicyEngine{}~\tPolicyEngine{}.

\subsection[Self-contained Honeypots]{\cHoneypots~Self-contained Honeypots}
\label{sec:isolated-honeypots}
\label{met:honeypot}

This work distinguishes between deception techniques 
that are intertwined with an application
and self-contained, sometimes isolated honeypot software 
that typically emulates services or protocols~\cite{Nawrocki2016:SurveyHoneypotSoftware}.
If honeypots are provided as container images~\citecloudbasedhoneypots{},
they can easily be added alongside application containers in the same pod,
or, for better isolation and to avoid restarting the application, as a new pod.
A \mKubernetesOperator{}~\tKubernetesOperator{}
or \mPolicyEngine{}~\tPolicyEngine{}
can automate and orchestrate the deployment of such honeypots.

\subsection[Reverse Proxies]{\cReverseProxies~Reverse Proxies}
\label{sec:reverse-proxies}
\label{met:reverse-proxy}

Placing reverse proxies in-front of applications is the most
commonly found method in research on application layer cyber deception%
~\citereverseproxysolutions{}. 
A reverse proxy accepts network requests on behalf of an application.
We can configure them to modify some HTTP response payloads
(e.g., to add decoy requests like the one in Listing~\ref{lst:network-decoy})
or have them respond to fake API endpoints that would not otherwise exist.
Proxies can also redirect, drop, or delay~\cite{Julian2002:DelayingTypeResponsesUse}
traffic in order to deceive.

In Kubernetes, one can either configure the ingress proxy at the perimeter,
or place a proxy in a \emph{sidecar container} inside the application pod,
which is then directly in-front of the application container.
The latter achieves better defense-in-depth,
as adversaries already acting from inside the cluster
will still connect to these proxies first when addressing services.
A \mKubernetesOperator{}~\tKubernetesOperator{}
or \mPolicyEngine{}~\tPolicyEngine{}
could be used to add these proxies to the deployment manifests of applications.
\mSecondaryContainers{}~\tsSecondaryContainers{}
and \mLifecycleHooks{}~\tLifecycleHooks{} could possibly
also change the configuration on disk before the proxy starts.

Sidecar containers are also commonly used by
\emph{service meshes}~\cite{Li2019:ServiceMeshChallenges},
which is a pattern that places a proxy in each pod,
so that all cluster traffic can be centrally monitored and controlled.
This pattern comes with a significant performance overhead~\citesidecaroverheadinresearch{},
which is why industry appears to be moving toward hybrid solutions
that aim to do much of the work in the kernel instead~\citesidecaroverheadinindustry{}.

\subsection{Container- and Pod-specific Methods}
\label{sec:container-pod-methods}

\subsubsection[Container File Modifications]{\cFileModifications~Container File Modifications}
\label{met:file-modifications}

This method requires a method that can modify files in a container,
such as a \mDeploymentAgent{}~\tDeploymentAgent{}.
We can then freely modify files in application containers or add honeytokens,
e.g., at well-known locations such as \texttt{\tildecentered{}/.ssh/id\_rsa}.

Theoretically, we could also modify some application artifacts directly.
If an application is written in an interpreted language such as Python or JavaScript,
we could patch the source code directly in the container.
Somewhat more complex and risky, but still possible,
would be to patch shared libraries or even binaries%
\cite{Araujo2014:PatchesHoneyPatchesLightweight}.

With this method, one often wants to patch files before the application starts.
This can be done using a wrapper script approach:
Let a \mKubernetesOperator{}~\tKubernetesOperator{}
override the container entrypoint (see \mLifecycleHooks{}~\tLifecycleHooks{})
to that of a \mDeploymentAgent{}~\tDeploymentAgent{} (the wrapper),
which will then patch the application before it starts.
The wrapper agent is added via \mVolumeMounts{}~\tVolumeMounts{}
or with an \emph{init container} (see \mSecondaryContainers{}~\tSecondaryContainers{}).

\subsubsection[Runtime Instrumentation]{\cRuntimeInstrumentation~Runtime Instrumentation}
\label{met:runtime-instrumentation}

Managed language runtimes, such as the Java Virtual Machine (JVM)
or the .NET Common Language Runtime (CLR)
provide the means to attach \emph{agents} or \emph{profilers}
that can inspect and control the execution of applications at runtime.
The JVM provides the JVM Tool Interface (JVM~TI)~\cite{Oracle2013:JVMTMTool}
and .NET provides the .NET CLR Profiling API~\cite{Microsoft2021:NETCLRProfiling}.
By hooking socket connection handlers
or injecting code that places honeytokens in the file system,
one can enrich an application with advanced cyber deception scenarios.

Runtime agents are typically attached by passing their path as an argument
to the runtime process~\cite{Oracle2013:JVMTMTool},
which requires overriding the entrypoint with \mLifecycleHooks{}~\tLifecycleHooks{}.

\subsubsection[LD\_PRELOAD Trick]{\cLdPreload~\texttt{LD\_PRELOAD} Trick}
\label{met:ld-preload}

By specifying the path of a shared library
in the \texttt{LD\_PRELOAD} environment variable or
placing the library in \texttt{/etc/ld.so.preload},
one instructs the dynamic linker, in Linux, to first load that library before any others.
Similar methods exist for other operating systems%
~\cite{Myers2004:InterceptingArbitraryFunctions}.
This method can be used to override or wrap the implementation of typical shared libraries,
such as libc, which powers many modern language runtimes. 
By hooking low-level functions such as \texttt{send} or \texttt{read},
one can modify network packets or file operations of an application,
also for cyber deception purposes~\cite{Kern2024:InjectingSharedLibraries}.

In Kubernetes, one can set the \texttt{LD\_PRELOAD} environment variable
in the container manifest with a \mKubernetesOperator{}~\tKubernetesOperator{}
or \mPolicyEngine{}~\tPolicyEngine{}.
Also, \mVolumeMounts{}~\tVolumeMounts{} are required to provide the shared library.

\subsubsection[Tracing with ptrace]{\cPtrace~Tracing with \texttt{ptrace}}
\label{met:ptrace}

The \texttt{ptrace} system call can, in Linux, be issued to ``trace'' another process.
This means that the \emph{tracer}
can inspect and control the execution and memory of the \emph{tracee}.
While this method is typically used by debuggers%
~\cite{Holl2018:KernelAssistedDebuggingLinux},
it is also rarely used to implement deception strategies%
~\cite{Araujo2020:ImprovingCybersecurityHygiene},
although, this then needs to be coded at the machine code level,
since that is the interface provided by \texttt{ptrace}.

\subsection{Kernel- and Network-specific Methods}
\label{sec:kernel-network-methods}

\subsubsection[Monitoring with eBPF]{\cEbpfMonitoring~Monitoring with eBPF}
\label{met:ebpf-monitoring}

The extended Berkeley Packet Filter (eBPF) is a Linux kernel technology
that makes the kernel programmable~\cite{Rice2022:WhatEBPF}.
It allows one to write small programs that hook into
system calls, network events, kernel and user probes, and more.
This technology is widely used to gain observability into system and network events%
~\cite{Soldani2023:EBPFNewApproach,Liu2020:ProtocolindependentContainerNetwork},
which makes it a natural fit for building captors.

In Kubernetes, one can use tools like Falco~\cite{TheFalcoAuthors:Falco}
or Tetragon~\cite{TheTetragonAuthors:Tetragon} to conveniently setup eBPF-based monitoring rules,
e.g., to alert when a honeytoken (placed by a \mDeploymentAgent{}~\tDeploymentAgent{})
is later opened by any process.

\subsubsection[Routing with Netfilter]{\cNetfilterRouting~Routing with Netfilter}
\label{met:netfilter}

Netfilter is a Linux kernel framework that is responsible
for controlling the flow of network packets.
Netfilter (and its ``frontend'' iptables) often assist in realizing
honeypot systems~\cite{%
  Franco2021:SurveyHoneypotsHoneynets, 
  Nawrocki2016:SurveyHoneypotSoftware, 
  Acosta2020:CybersecurityDeceptionExperimentation, 
  Provos2007:VirtualHoneypotsBotnet, 
  Borders2007:OpenFireUsingDeception, 
  Reti2023:HoneyInfiltratorInjecting 
}.
To implement application layer cyber deception techniques,
one could follow an approach similar to the one proposed by
\citeauthor{Reti2023:HoneyInfiltratorInjecting}~\cite{Reti2023:HoneyInfiltratorInjecting},
where Netfilter is used to redirect packets to a user space process that modifies payloads,
before they are sent to the their original destination.

\subsubsection[Routing with eBPF]{\cEbpfRouting~Routing with eBPF}
\label{met:ebpf-routing}

eBPF can also be used to redirect packets%
~\cite{
  Tu2017:BuildingExtensibleOpen,
  Miano2018:CreatingComplexNetwork,
  DeLosSantos2023:EBPFNetworkingTechniques}.
Similar to the \mNetfilterRouting{}~\tNetfilterRouting{} method,
we could also use this method
to redirect packets to a user space process, modify their payloads in user space,
and then send them back to their original destination.
However, there are significant challenges to using eBPF~\cite{%
  Miano2018:CreatingComplexNetwork,
  Vieira2020:FastPacketProcessing,
  Tran2020:SocketOptionsFully},
which probably explains why we have not found any works that
use it to implement application (rather than network) layer cyber deception.
Nevertheless, solutions that facilitate the eBPF developer experience%
~\cite{Miano2021:FrameworkEBPFBasedNetwork}
seem promising for future research.

\subsubsection[File System Implementations]{\cFilesystemImplementations~File System Implementations}
\label{met:filesystem-implementations}

The Linux kernel allows to overlay, merge, and implement custom file systems
(even in user space, with FUSE~\cite{Szeredi:Libfuse}).
We can use this to place honeytokens in file systems, monitor file access,
and perform \cFileModifications{}~\tFileModifications{}.
This method is commonly used for malware detection
and cyber decepetion use cases~\cite{%
  vonderAssen2024:GuardFSFileSystem,
  vonderAssen2023:MTFSMovingTarget,
  Taylor2018:HiddenPlainSight,
  Kohlbrenner2017:POSTERHiddenPlain,
  Araujo2014:PatchesHoneyPatchesLightweight
}

\subsection{Container Platform-specific Methods}
\label{sec:kubernetes-methods}

\subsubsection[Volume Mounts]{\cVolumeMounts~Volume Mounts}
\label{met:volume-mounts}

If one cannot make \mFileModifications{}~\tFileModifications{} directly,
an alternative option is to mount files and directories into application containers.
This requires modifying the deployment manifest with
a \mKubernetesOperator{}~\tKubernetesOperator{} or \mPolicyEngine{}~\tPolicyEngine{},
as well as a restart of the affected pod for the initial setup.

This method offers a convenient way to add honeytokens to container file systems.
If the volume is not mounted read-only, these files can also be modified at runtime.
Despite its simplicity, we are not aware of any research that has reported using this method.

\subsubsection[Policy Management]{\cPolicyEngine~Policy Management}
\label{met:policy-management}

Kubernetes provides policy management solutions~\cite{Salier2022:KubernetesPolicyManagement}
such as Kyverno~\cite{TheKyvernoAuthors:Kyverno},
that give cluster administrators a convenient way to specify and enforce cluster-wide policies.
Among many other things,
a policy can enforce that new pods have \mVolumeMounts{}~\tVolumeMounts{} with honeytokens,
that sidecar containers with \mReverseProxies{}~\tReverseProxies{} exist,
that the \mLdPreload{}~\tLdPreload{} environment variable is set,
or that an \mEbpfMonitoring{}~\tEbpfMonitoring{}-based tracing policy exists.
To enforce network policies, a compatible \mCni{}~\tCni{} is also needed.

\subsubsection[CNI Plugin]{\cCni~CNI Plugin}
\label{met:cni}

A Container Network Interface (CNI) plugin such as Cilium~\cite{TheCiliumAuthors:Cilium}
implements the Kubernetes network model.
This layer could potentially implement network layer cyber deception,
though it may be challenging to lift that up to the application layer.
This may be why we have not yet found any works on that.

\begin{table*}[t]
  \begin{threeparttable}
    \footnotesize
    \centering
    \caption{Overview of technical methods to achieve application layer cyber deception}
    \label{tab:overview}
    \input{./figures/table.tex}
    \begin{tablenotes}
      \footnotesize
      \item \textbf{Description:} The table shows properties (\S\ref{sec:properties})
      of each method (\S\ref{sec:methods}). Appendix~\ref{sec:appendix:rubric}
      lists questions that we asked for each individual cell.
      \vspace{0.1cm}
      \item
      \textbf{D-C-S Type}
      fits either \FullMark~excellently, \HalfMark~partially, or \EmptyMark~not at all.
      \textbf{Use Cases}
      are supported \FullMark~excellently, \HalfMark~partially, or \EmptyMark~not at all.
      \textbf{Deployment Mode}
      is either \DeplBuiltIn~built-in (at the design phase),
      \DeplInFrontOf~in-front of (e.g., proxies, hooks),
      or \DeplAddedTo~added-to (e.g., files, containers).
      \textbf{Layer of Deception}
      is either network~(\LayerNetwork),
      system~(\LayerSystem),
      application~(\LayerApplication),
      or data~(\LayerData).
      \textbf{Application Dimension}
      is either source code~(\DimSource),
      runtime~(\DimRuntime),
      shared libraries~(\DimLibraries),
      container~(\DimContainer),
      or kernel~(\DimKernel).
      \textbf{Software Dependencies}
      are either Kubernetes~(\ResKubernetes),
      an OCI-compliant container platform (\ResOciCompliant),
      a service mesh~(\ResServiceMesh),
      a policy management solution~(\ResPolicyEngine),
      a CNI plugin~(\ResCniPlugin),
      an application relying on the C standard library~(\ResLibc),
      an application running in a managed language runtime~(\ResManagedRuntime),
      an eBPF-enabled kernel~(\ResEbpf),
      a system with Netfilter~(\ResNetfilter), or no dependencies. 
      \textbf{Operational properties}
      and \textbf{efficacy} follow an absolute category rating
      (Appendix~\ref{sec:appendix:rubric}):
      \ExcellentMark~excellent, \GoodMark~good, \FairMark~fair, \PoorMark~poor, and \BadMark~bad.
      \vspace{0.1cm}
      \item[1] We consider honeypots as being added-to the same cluster,
      and not as stand-alone to a production system.
      \item[2] This method could also be considered part of the system layer
      when it is installed on the host, or when it needs to interact with the kernel.
      \item[3] This method could also be considered part of the application layer
      when used to interact with application artifacts.
      \item[4] Typical policies want to directly modify workload artifacts and
      be transparent to the system operator, but exceptions may be technically feasible.
      \item[5] Only ephemeral containers can be added and patched
      without restarting the application. Init and sidecar only run after the pod is recreated.
      \item[6] Except for modifying (binary) application artifacts that are not hot-reloaded,
      which would require an application restart then.
      \item[7] Assuming that the runtime supports live agent injection.
      Some technologies such as JVM TI~\cite{Oracle2013:JVMTMTool}
      do support that for certain environments.
    \end{tablenotes}
  \end{threeparttable}
\end{table*}

\subsection{Methods with Developer Interaction}
\label{sec:developer-interaction-methods}

\subsubsection[Deception Software Framework]{\cDeceptionFramework~Deception Software Framework}
\label{met:deception-framework}

Instead of costly function hooking and patching,
one could instead embed (standardized) interfaces within applications at build time
(in the ``Develop'' phase~\cite{Krieger2022:CloudNativeSecurity})
that allow dynamic attachment and configuration of deception scenarios.
Such a tool is best provided as an easy-to-integrate software framework
that provides standard use cases (e.g., honeytokens, fake API endpoints)
out-of-the-box and also allows for further customization.
Despite the prevalence of the term ``deception framework''
in the literature~\citemalwaredeceptionframeworks{},
its use appears to refer mostly to theoretical concepts,
but software frameworks or respective efforts do not seem
to appear in the scientific literature yet.

\subsubsection[Build Process Intervention]{\cBuildProcess~Build Process Intervention}
\label{met:build-process}

Access to an application's source code may not always be possible,
but when one can at least control the build pipeline
(in the ``Distribute'' phase~\cite{Krieger2022:CloudNativeSecurity}),
container- and pod-specific methods~(\S\ref{sec:container-pod-methods})
as well as \mHoneypots{}~\tHoneypots{} or \mReverseProxies{}~\tReverseProxies{}
can already be incorporated at build time.

%% file: figures/overview.tex
{\scalefont{1.0}\begin{tikzpicture}[scale=0.66,on grid]
  \def\podoffset{.25}%
  \def\arrowoffset{.2}%
  \def\yoffset{1}%
  \def\boxheight{1}%
  \def\boxheightsm{0.8}%
  \def\boxwidth{1.1}%
  \def\boxwidthsm{0.8}%
  \def\methodscale{0.75}%
  \def\honeypotshrink{0}%

  \coordinate (ClusterTL) at (-14-\podoffset-.5,4.25+.125);
  \coordinate (ClusterTR) at (8.75,4.25+.125);
  \coordinate (ClusterBR) at (8.75,-9.5-.125); 
  \coordinate (NodeTR) at (4,2);
  \coordinate (NodeBL) at (-13.5-\podoffset-.5,-9-.125); 

  \coordinate (PipelineTL) at (8.75+1,4.25+.125);
  \coordinate (PipelineBR) at (14.5+.125,-9.5-.125);

  \coordinate (CicdTL) at (8.75+1+0.5,2);
  \coordinate (CicdBR) at ($(CicdTL)+(3.5*\boxwidth,-3*\boxheight)$);

  \coordinate (CicdInnerTL) at ($(CicdTL)+(0.5,-0.5)$);
  \coordinate (CicdInnerBR) at ($(CicdBR)+(-0.5,0.5)$);

  \coordinate (FrameworkTL) at (8.75+1+0.5,-2.75);
  \coordinate (FrameworkBR) at ($(FrameworkTL)+(3.5*\boxwidth,-3*\boxheight)$);

  \coordinate (FrameworkInnerTL) at ($(FrameworkTL)+(0.5,-0.5)$);
  \coordinate (FrameworkInnerBR) at ($(FrameworkBR)+(-0.5,0.5)$);

  \coordinate (KProxyTR) at (-0.25+\podoffset,2+0.33*\boxheight);
  \coordinate (KProxyBL) at ($(KProxyTR)+(-3.5,-\boxheight)$);

  \coordinate (IngressTL) at (-15.5-.5,4);
  \coordinate (IngressBR) at ($(IngressTL)+(2.5,-\boxheight)$);

  \coordinate (FilterBL) at ($(IngressBR)+(1.25,0)$);
  \coordinate (FilterTR) at ($(FilterBL)+(3*\boxwidthsm,\boxheight)$);

  \coordinate (StickT) at (-15.5-.5, 0);
  \coordinate (DevT) at (12.25,-4-\podoffset-4*\boxheight);

  \coordinate (ServiceTL) at (-8+\podoffset,4);
  \coordinate (ServiceBR) at ($(ServiceTL)+(2.5,-\boxheight)$);
  \coordinate (ServiceBgTL) at ($(ServiceTL)+(\podoffset,-\podoffset)$);
  \coordinate (ServiceBgBR) at ($(ServiceBR)+(\podoffset,-\podoffset)$);

  \coordinate (CtrTR) at (2+\podoffset,2-2.5*\yoffset-\podoffset);
  \coordinate (CtrBL) at (-7.25+\podoffset-.5,2-3.5*\yoffset-\podoffset-3*\boxheight-0.5-2*\boxheightsm);

  \coordinate (PodTR) at (2.5+\podoffset,2-\yoffset-\podoffset);
  \coordinate (PodBL) at ($(CtrBL)+(-.5-2.25,-.5)$);  
  \coordinate (PodBgTR) at ($(PodTR)+(\podoffset,\podoffset)$);
  \coordinate (PodBgBL) at ($(PodBL)+(\podoffset,\podoffset)$);

  \coordinate (PodSecBR) at ($(PodBL)+(-0.5,0)$);
  \coordinate (PodSecTL) at ($(PodSecBR)+(-2-\boxwidthsm,2+0.5+\yoffset+3*\boxheight+2*\boxheightsm)$);

  \coordinate (CtrSecTL) at ($(PodSecTL)+(0.5,-1.5)$);
  \coordinate (CtrSecBR) at ($(PodSecBR)+(-0.5,0.5+\honeypotshrink)$);

  \coordinate (HoneypotTL) at ($(CtrSecTL)+(0.5,-0.5)$);
  \coordinate (HoneypotBR) at ($(CtrSecBR)+(-0.5,0.5)$);

  \coordinate (SecondaryBL) at ($(PodBL)+(0.5,0.5)$);
  \coordinate (SecondaryTR) at ($(SecondaryBL)+(1+\boxwidthsm,2*\boxheightsm)$);

  \coordinate (SidecarBR) at ($(SecondaryTR)+(0,0.5)$);
  \coordinate (SidecarTL) at ($(CtrTR)+(-11.5-.5,0)$); 

  \coordinate (EnvoyTL) at ($(SidecarTL)+(0.5,-0.5)$);
  \coordinate (EnvoyBR) at ($(SidecarBR)+(-0.5,0.5)$);

  \coordinate (AppTL) at ($(CtrTR)+(.5-4.25-5-.5,-.5)$);
  \coordinate (AppBR) at ($(AppTL)+(3.6,-\boxheight)$);
  \coordinate (RuntimeTL) at ($(AppTL)+(0,-\boxheight)$);
  \coordinate (RuntimeBR) at ($(RuntimeTL)+(3.6,-\boxheight)$);
  \coordinate (LibsTL) at ($(RuntimeTL)+(0,-\boxheight)$);
  \coordinate (LibsBR) at ($(LibsTL)+(3.6,-\boxheight)$);

  \coordinate (AgentTL) at ($(AppTL)+(5+2.4+.5,0)$);
  \coordinate (AgentBR) at ($(AgentTL)+(\boxwidthsm,-3*\boxheight)$);

  \coordinate (PtraceTL) at ($(LibsTL)+(0,-0.5-\boxheight)$);
  \coordinate (PtraceBR) at ($(PtraceTL)+(2.5*\boxwidthsm,-2*\boxheightsm)$);

  \coordinate (FilesTR) at ($(PtraceTL)+(0.5+2.5*\boxwidthsm,0)$);
  \coordinate (FilesBL) at ($(FilesTR)+(4*\boxwidthsm,-2*\boxheightsm)$);

  \coordinate (MountsTL) at ($(FilesTR)+(0.5+4*\boxwidthsm,0)$);
  \coordinate (MountsBR) at ($(MountsTL)+(3.125*\boxwidthsm,-2*\boxheightsm)$);

  \coordinate (InstrumentTR) at ($(RuntimeTL)+(6.6+.5,\boxheight)$);
  \coordinate (InstrumentBL) at ($(InstrumentTR)+(-3.5,-2*\boxheight)$);

  \coordinate (PreloadTL) at ($(LibsTL)+(6.6+.5,0)$);
  \coordinate (PreloadBR) at ($(PreloadTL)+(-3.5,-\boxheight)$);

  \coordinate (LifecycleTL) at ($(CtrTR)+(-8.75,\boxheightsm+0.35)$);
  \coordinate (LifecycleBR) at ($(LifecycleTL)+(3.6,-\boxheightsm)$);

  \coordinate (KernelTR) at (2.5+\podoffset,-7.75);
  \coordinate (KernelBL) at (-13.5+0.2+\podoffset,-7.75-\boxheight);

  \coordinate (EbpfLeftTR) at ($(KernelTR)+(-15.8-.5,0)$);
  \coordinate (EbpfLeftBL) at ($(EbpfLeftTR)+(3.5*\boxwidthsm,-\boxheight)$);

  \coordinate (EbpfLeftSecondTL) at ($(EbpfLeftBL)+(0,\boxheight)$);
  \coordinate (EbpfLeftSecondBR) at ($(EbpfLeftSecondTL)+(3.5*\boxwidthsm,-\boxheight)$);

  \coordinate (EbpfRightTR) at ($(KernelTR)+(0,0)$);
  \coordinate (EbpfRightBL) at ($(EbpfRightTR)+(-3.75*\boxwidthsm-1.05,-\boxheight)$);

  \coordinate (ControlTL) at (5,-2.5-\podoffset);
  \coordinate (ControlBR) at (5+2*\boxwidth+1,-4-\podoffset-4*\boxheight);

  \coordinate (OperatorTL) at (5.5,2);
  \coordinate (OperatorBR) at (5.5+2*\boxwidth,2-3*\boxheightsm);

  \coordinate (ApiServerTL) at ($(ControlTL)+(0.5,-0.5)$);
  \coordinate (ApiServerBR) at ($(ApiServerTL)+(2*\boxwidth,-2*\boxheight)$);

  \coordinate (PolicyBR) at ($(ControlBR)+(-0.5,0.5)$);
  \coordinate (PolicyTL) at ($(PolicyBR)+(-2*\boxwidth,2*\boxheightsm)$);

  \coordinate (NetworkTL) at ($(NodeTR)+(-.5*\boxwidth,-.5)$);
  \coordinate (NetworkBR) at ($(NetworkTL)+(\boxwidth,-9.75+4*\boxheightsm)$);

  \coordinate (CniTL) at ($(NetworkBR)+(0,-4*\boxheightsm)$);
  \coordinate (CniBR) at ($(CniTL)+(-\boxwidth,4*\boxheightsm)$);

  \draw[dotted] (ClusterTL) rectangle (ClusterBR){};
  \node[below left,black!40,text width=2cm,align=right,scale=\methodscale] at (ClusterTR) {\textbf{Cluster}\\Domain};

  \draw[dotted] (PipelineTL) rectangle (PipelineBR){};
  \node[below right,black!40,text width=2cm,align=left,scale=\methodscale] at (PipelineTL) {\textbf{Developer}\\Domain};

  \draw[dashed] (CicdTL) rectangle (CicdBR){};
  \node[above right] at (CicdTL) {CI/CD};
  \draw[OutlineG,fill=FillG,thick] (CicdInnerTL) rectangle (CicdInnerBR) node[ColorG,midway,align=center,scale=\methodscale] {Build\\Process\\Intervent.};
  \node[anchor=center] at ($(CicdInnerTL)+(1.25*\boxwidth,0)$) {\mBuildProcess[1]};

  \draw[dashed] (FrameworkTL) rectangle (FrameworkBR){};
  \node[above right] at (FrameworkTL) {Source Code};
  \draw[OutlineG,fill=FillG,thick] (FrameworkInnerTL) rectangle (FrameworkInnerBR) node[ColorG,midway,align=center,scale=\methodscale] {Deception\\Software\\Framework};
  \node[anchor=center] at ($(FrameworkInnerTL)+(1.25*\boxwidth,0)$) {\mDeceptionFramework[1]};

  \draw[pattern=north west lines,pattern color=black!10] (NodeTR) rectangle (NodeBL){};
  \node[above left] at (NodeTR) {Node};

  \draw[fill=white] (PodBgTR) rectangle (PodBgBL){};
  \draw[fill=white] (PodTR)[fill=white] rectangle (PodBL){};
  \node[above left] at (PodBgTR) {Pods};

  \draw[fill=white] (PodSecTL)[fill=white] rectangle (PodSecBR){};
  \node[above right] at (PodSecTL) {Pod};

  \draw[dashed] (CtrSecTL) rectangle (CtrSecBR){};
  \node[above] at ($(CtrSecTL)+(0.5+0.5*\boxwidthsm,0)$) {Ctr.};

  \draw[OutlineB,fill=FillB,thick] (HoneypotTL) rectangle (HoneypotBR) node[ColorB,midway,align=center,scale=\methodscale,rotate=90] {Honeypot};
  \node[anchor=center] at ($(HoneypotTL)+(0.5*\boxwidthsm,0)$) {\mHoneypots[1]};

  \draw[dashed] (CtrTR) rectangle (CtrBL){};
  \node[above left] at (CtrTR) {App Container};

  \draw[OutlineA,fill=FillA,dashed,thick] (SecondaryBL) rectangle (SecondaryTR) node[ColorA,midway,align=center,scale=\methodscale] {Second.\\Ctr.};
  \node[anchor=center] at ($(SecondaryTR)+(-0.5-0.5*\boxwidthsm,0)$) {\mSecondaryContainers[1]};

  \draw[dashed] (SidecarTL) rectangle (SidecarBR){};
  \node[above] at ($(SidecarTL)+(0.5+0.5*\boxwidthsm,0)$) {Sidecar};

  \draw[OutlineC,fill=FillC,thick] (EnvoyTL) rectangle (EnvoyBR) node[ColorC,midway,align=center,scale=\methodscale,rotate=90] {Proxy};
  \node[anchor=center] at ($(EnvoyTL)+(0.5*\boxwidthsm,0)$) {\mReverseProxies[1]};

  \draw[fill=SteelBlue1!50] (AppTL) rectangle (AppBR) node[midway,align=center]{Application};
  \draw (RuntimeTL) rectangle (RuntimeBR) node[midway,align=center]{Runtime};
  \draw (LibsTL) rectangle (LibsBR) node[midway,align=center]{Shared Libs.};

  \draw[OutlineA,fill=FillA,thick] (AgentTL) rectangle (AgentBR) node[ColorA,midway,align=center,scale=\methodscale,rotate=90] {Agent};
  \node[anchor=center] at ($(AgentTL)+(0.5*\boxwidthsm,0)$) {\mDeploymentAgent[1]};

  \draw[OutlineF,fill=FillF,thick] (MountsTL) rectangle (MountsBR) node[ColorF,midway,align=center,scale=\methodscale] {Volume\\Mounts};
  \node[anchor=center] at ($(MountsTL)+(1.5625*\boxwidthsm,0)$) {\mVolumeMounts[1]};

  \draw[OutlineD,fill=FillD,thick] (FilesTR) rectangle (FilesBL) node[ColorD,midway,align=center,scale=\methodscale] {Container File\\Modifications};
  \node[anchor=center] at ($(FilesTR)+(2*\boxwidthsm,0)$) {\mFileModifications[1]};

  \draw[OutlineD,fill=FillD,thick] (InstrumentTR) rectangle (InstrumentBL) node[ColorD,midway,align=center,scale=\methodscale] {Runtime\\Instrumentation};
  \node[anchor=center] at ($(InstrumentTR)+(0,-\boxheight)$) {\mRuntimeInstrumentation[1]};

  \draw[OutlineD,fill=FillD,thick] (PreloadTL) rectangle (PreloadBR) node[ColorD,midway,align=center,scale=\methodscale] {\texttt{LD\_PRELOAD}};
  \node[anchor=center] at ($(PreloadTL)+(0,-0.5*\boxheight)$) {\mLdPreload[1]};

  \draw[OutlineD,fill=FillD,thick] (PtraceTL) rectangle (PtraceBR) node[ColorD,midway,align=center,scale=\methodscale] {\texttt{ptrace}};
  \node[anchor=center] at ($(PtraceTL)+(1.25*\boxwidthsm,0)$) {\mPtrace[1]};

  \draw[OutlineA,fill=FillA,thick] (LifecycleTL) rectangle (LifecycleBR) node[ColorA,midway,align=center,scale=\methodscale] {Lifecycle Hooks};
  \node[anchor=center] at ($(LifecycleTL)+(0,-0.5*\boxheightsm)$) {\mLifecycleHooks[1]};

  \draw[dashed] (ControlTL) rectangle (ControlBR) {};
  \node[below] at ($(ControlBR)+(-1.5*\boxwidth,0)$) {Control Plane};

  \draw (ApiServerTL) rectangle (ApiServerBR) node[midway,align=center]{API\\Server};
  \draw[OutlineF,fill=FillF,thick] (PolicyTL) rectangle (PolicyBR) node[ColorF,midway,align=center,scale=\methodscale] {Policy\\Mngmt.};
  \node[anchor=center] at ($(PolicyTL)+(1*\boxwidth,0)$) {\mPolicyEngine[1]};

  \draw[OutlineA,fill=FillA,thick] (OperatorTL) rectangle (OperatorBR) node[ColorA,midway,align=center,scale=\methodscale] {Kubernetes\\Operator\\and API};
  \node[anchor=center] at ($(OperatorTL)+(1*\boxwidth,0)$) {\mKubernetesOperator[1]};

  \draw[fill=yellow!20,dashed] (NetworkTL) rectangle (NetworkBR) node[midway,align=center,rotate=90] {Network};

  \draw[OutlineF,fill=FillF,thick] (CniTL) rectangle (CniBR) node[ColorF,midway,align=center,scale=\methodscale,rotate=90] {CNI Plugin};
  \node[anchor=center] at ($(CniBR)+(0.5*\boxwidth,0)$) {\mCni[1]};

  \draw[fill=white] (KernelTR) rectangle (KernelBL) node[midway,align=center]{Host Kernel};

  \draw[OutlineE,fill=FillE,thick] (EbpfLeftTR) rectangle (EbpfLeftBL) node[ColorE,midway,align=center,scale=\methodscale] {eBPF};
  \draw[OutlineE,fill=FillE,thick] (EbpfLeftSecondTL) rectangle (EbpfLeftSecondBR) node[ColorE,midway,align=center,scale=\methodscale] {Netfilter};
  \draw[OutlineE,fill=FillE,thick] (EbpfRightTR) rectangle (EbpfRightBL) node[ColorE,midway,align=center,scale=\methodscale] {Filesystem Impl.};

  \node[anchor=center] at ($(EbpfLeftBL)+(-3.5*\boxwidthsm,0.85*\boxheight)$) {\mEbpfMonitoring[1]};
  \node[anchor=center] at ($(EbpfLeftBL)+(-3.5*\boxwidthsm,0.15*\boxheight)$) {\mEbpfRouting[1]};
  \node[anchor=center] at ($(EbpfLeftSecondBR)+(0,0.5*\boxheight)$) {\mNetfilterRouting[1]};
  \node[anchor=center] at ($(EbpfRightBL)+(0,0.5*\boxheight)$) {\mFilesystemImplementations[1]};

  \draw[fill=white] (KProxyTR) rectangle (KProxyBL) node[midway,align=center]{kube-proxy};
  \draw[fill=white] (IngressTL) rectangle (IngressBR) node[midway,align=center]{Ingress};

  \draw[fill=white] (ServiceBgTL) rectangle (ServiceBgBR){};
  \draw[fill=white] (ServiceTL) rectangle (ServiceBR) node[midway,align=center]{Services};

  \draw[-{Stealth[scale=1.25]},very thick] ($(IngressBR)+(\arrowoffset,0.5*\boxheight)$) -- ($(ServiceTL)+(-\arrowoffset,-0.5*\boxheight)$);
  \draw[-{Stealth[scale=1.25]},very thick] ($(ServiceBR)+(\podoffset+\arrowoffset,0.5*\boxheight)$)
  -- ($(ServiceBR)+(3.25,0.5*\boxheight)$) to[out=0,in=90] ($(KProxyTR)+(-.5*3.5,\arrowoffset)$);
  \draw[-{Stealth[scale=1.25]},very thick] ($(KProxyBL)+(3.5*0.5,-\arrowoffset)$) to[out=-90,in=60] ($(CtrTR)+(-4.5,\arrowoffset)$);
  \draw[-{Stealth[scale=1.25]},very thick] ($(KProxyBL)+(-\arrowoffset,0.33*\boxheight)$)
  -- ($(KProxyBL)+(-\arrowoffset-4,0.33*\boxheight)$) to[out=180,in=90] ($(SidecarTL)+(0.5+0.5*\boxwidthsm,0.75)$);
  \draw[-{Stealth[scale=1.25]},very thick] ($(KProxyBL)+(-\arrowoffset,0.33*\boxheight)$)
  -- ($(KProxyBL)+(-\arrowoffset-7.25,0.33*\boxheight)$) to[out=180,in=90] ($(CtrSecTL)+(0.5+0.5*\boxwidthsm,0.75)$);
  \draw[-{Stealth[scale=1.25]},very thick] ($(StickT)+(-.15,.5)$) -- ($(StickT)+(-.15,1)$) to[out=90,in=240] ($(IngressTL)+(.5,-\boxheight-\arrowoffset)$);

  \draw[-{Stealth[scale=1.25]},very thick] ($(OperatorBR)+(-\boxwidth,-\arrowoffset)$) -- ($(ControlTL)+(\boxwidth+0.5,\arrowoffset)$);

  \draw[-{Stealth[scale=1.25]},very thick] ($(FrameworkTL)+(3.5*\boxwidth-\arrowoffset,\arrowoffset)$) -- ($(CicdBR)+(-\arrowoffset,-\arrowoffset)$);
  \draw[-{Stealth[scale=1.25]},very thick] ($(DevT)+(1,1)$) to[out=20,in=-90] ($(FrameworkBR)+(-\arrowoffset,-\arrowoffset)$);

  \draw[-{Stealth[scale=1.25]},thick,solid,black!40] ($(PipelineTL)+(-\arrowoffset,-.55)$) to[out=180,in=0] ($(ClusterTR)+(\arrowoffset,-.55)$);

  \node[alice,anchor=north,scale=1.75] at (StickT){};
  \node[anchor=center] at ($(StickT)+(0,-2)$){Users};

  \node[bob,anchor=north,scale=1.75] at ($(DevT)+(0,-0.425+2)$){};
  \node[anchor=center] at ($(DevT)+(0,-0.425)$){Developers};

  \draw[OutlineC,fill=FillC,thick] (FilterBL) rectangle (FilterTR) node[ColorC,midway,align=center,scale=\methodscale] {Proxy};
  \node[anchor=center] at ($(FilterTR)+(-3*\boxwidthsm,-0.5*\boxheight)$) {\mReverseProxies[1]};
\end{tikzpicture}}

%% file: figures/table.tex
\newcommand\RHead[1]{\rotatebox{90}{\varwidth{\linewidth}#1\endvarwidth}}
\newcommand{\methodheader}[1]{{\hypersetup{hidelinks}\RHead{#1}}}

\NewKeyValTable{Methods}{
  prop: align=X, head={Property};
  DeploymentAgent: align=c, head=\methodheader{\hyperref[met:deployment-agent]{Deployment Agent}};
  KubernetesOperator: align=c, head=\methodheader{\hyperref[met:kubernetes-operator]{K8s Operator}};
  SecondaryContainers: align=c, head=\methodheader{\hyperref[met:secondary-container]{Secondary Ctrs.}};
  LifecycleHooks: align=c, head=\methodheader{\hyperref[met:lifecycle-hooks]{Ctr. Lifecycle Hooks}};
  Honeypots: align=c, head=\methodheader{\hyperref[met:honeypot]{Self-cont. Honeypots}};
  ReverseProxies: align=c, head=\methodheader{\hyperref[met:reverse-proxy]{Reverse Proxies}};
  FileModifications: align=c, head=\methodheader{\hyperref[met:file-modifications]{Ctr. File Mods.}};
  RuntimeInstrumentation: align=c, head=\methodheader{\hyperref[met:runtime-instrumentation]{Runtime Instrument.}};
  LdPreload: align=c, head=\methodheader{\hyperref[met:ld-preload]{\texttt{LD\_PRELOAD} Trick}};
  Ptrace: align=c, head=\methodheader{\hyperref[met:ptrace]{Tracing w/ \texttt{ptrace}}};
  EbpfMonitoring: align=c, head=\methodheader{\hyperref[met:ebpf-monitoring]{Monitoring w/ eBPF}};
  NetfilterRouting: align=c, head=\methodheader{\hyperref[met:netfilter]{Routing w/ Netfilter}};
  EbpfRouting: align=c, head=\methodheader{\hyperref[met:ebpf-routing]{Routing w/ eBPF}};
  FilesystemImplementations: align=c, head=\methodheader{\hyperref[met:filesystem-implementations]{File System Impl.}};
  VolumeMounts: align=c, head=\methodheader{\hyperref[met:volume-mounts]{Volume Mounts}};
  PolicyEngine: align=c, head=\methodheader{\hyperref[met:policy-management]{Policy Management}};
  Cni: align=c, head=\methodheader{\hyperref[met:cni]{CNI Plugin}};
  DeceptionFramework: align=c, head=\methodheader{\hyperref[met:deception-framework]{Decept. Framework}};
  BuildProcess: align=c, head=\methodheader{\hyperref[met:build-process]{Build Proc. Intervent.}};
}[colgroups={
      GroupSupportingMethods: span=DeploymentAgent+KubernetesOperator+SecondaryContainers+LifecycleHooks;
      GroupHoneypots: span=Honeypots;
      GroupReverseProxies: span=ReverseProxies;
      GroupContainersPods: span=FileModifications+RuntimeInstrumentation+LdPreload+Ptrace;
      GroupKernelNetwork: span=EbpfMonitoring+NetfilterRouting+EbpfRouting+FilesystemImplementations;
      GroupContainerPlatform: span=VolumeMounts+PolicyEngine+Cni;
      GroupDevInteraction: span=DeceptionFramework+BuildProcess;
    }]

\setlength{\tabcolsep}{3.1pt}

\begin{KeyValTable}[%
    backend=tabularx,
    width=\textwidth,
    shape=onepage,
    nobg=true,
    valign=c,
  ]{Methods}
  \Row[below=.5ex]{
    GroupSupportingMethods={\hypersetup{hidelinks}\color{OutlineA} \hyperref[sec:supporting-methods]{Supporting Methods}},
    GroupHoneypots=,
    GroupReverseProxies=,
    GroupContainersPods={\hypersetup{hidelinks}\color{OutlineD} \hyperref[sec:container-pod-methods]{Containers \& Pods}},
    GroupKernelNetwork={\hypersetup{hidelinks}\color{OutlineE} \hyperref[sec:kernel-network-methods]{Kernel \& Network}},
    GroupContainerPlatform={\hypersetup{hidelinks}\color{OutlineF} \hyperref[sec:kubernetes-methods]{Container Platform}},
    GroupDevInteraction={\hypersetup{hidelinks}\color{OutlineG} \hyperref[sec:developer-interaction-methods]{Dev. Inter.}},
  }
  \Row{
    DeploymentAgent=\scriptsize\mDeploymentAgent,
    KubernetesOperator=\scriptsize\mKubernetesOperator,
    SecondaryContainers=\scriptsize\mSecondaryContainers,
    LifecycleHooks=\scriptsize\mLifecycleHooks,
    Honeypots=\scriptsize\mHoneypots,
    ReverseProxies=\scriptsize\mReverseProxies,
    FileModifications=\scriptsize\mFileModifications,
    RuntimeInstrumentation=\scriptsize\mRuntimeInstrumentation,
    LdPreload=\scriptsize\mLdPreload,
    Ptrace=\scriptsize\mPtrace,
    EbpfMonitoring=\scriptsize\mEbpfMonitoring,
    NetfilterRouting=\scriptsize\mNetfilterRouting,
    EbpfRouting=\scriptsize\mEbpfRouting,
    FilesystemImplementations=\scriptsize\mFilesystemImplementations,
    VolumeMounts=\scriptsize\mVolumeMounts,
    PolicyEngine=\scriptsize\mPolicyEngine,
    Cni=\scriptsize\mCni,
    DeceptionFramework=\scriptsize\mDeceptionFramework,
    BuildProcess=\scriptsize\mBuildProcess,
  }
  \MidRule
  \Row{prop={\bfseries D-C-S Type}}
  \Row{prop={Decoy},
    DeploymentAgent=\EmptyMark,
    KubernetesOperator=\EmptyMark,
    SecondaryContainers=\HalfMark,
    LifecycleHooks=\EmptyMark,
    Honeypots=\FullMark,
    ReverseProxies=\FullMark,
    FileModifications=\FullMark,
    RuntimeInstrumentation=\FullMark,
    LdPreload=\FullMark,
    Ptrace=\FullMark,
    EbpfMonitoring=\EmptyMark,
    NetfilterRouting=\HalfMark,
    EbpfRouting=\HalfMark,
    FilesystemImplementations=\FullMark,
    VolumeMounts=\FullMark,
    PolicyEngine=\EmptyMark,
    Cni=\HalfMark,
    DeceptionFramework=\FullMark,
    BuildProcess=\HalfMark
  }
  \Row{prop={Captor},
    DeploymentAgent=\HalfMark,
    KubernetesOperator=\HalfMark,
    SecondaryContainers=\HalfMark,
    LifecycleHooks=\EmptyMark,
    Honeypots=\HalfMark,
    ReverseProxies=\HalfMark,
    FileModifications=\EmptyMark,
    RuntimeInstrumentation=\FullMark,
    LdPreload=\FullMark,
    Ptrace=\FullMark,
    EbpfMonitoring=\FullMark,
    NetfilterRouting=\HalfMark,
    EbpfRouting=\HalfMark,
    FilesystemImplementations=\HalfMark,
    VolumeMounts=\EmptyMark,
    PolicyEngine=\FullMark,
    Cni=\HalfMark,
    DeceptionFramework=\FullMark,
    BuildProcess=\HalfMark
  }
  \Row{prop={Support},
    DeploymentAgent=\FullMark,
    KubernetesOperator=\FullMark,
    SecondaryContainers=\FullMark,
    LifecycleHooks=\FullMark,
    Honeypots=\EmptyMark,
    ReverseProxies=\HalfMark,
    FileModifications=\HalfMark,
    RuntimeInstrumentation=\EmptyMark,
    LdPreload=\EmptyMark,
    Ptrace=\EmptyMark,
    EbpfMonitoring=\EmptyMark,
    NetfilterRouting=\FullMark,
    EbpfRouting=\FullMark,
    FilesystemImplementations=\HalfMark,
    VolumeMounts=\HalfMark,
    PolicyEngine=\FullMark,
    Cni=\FullMark,
    DeceptionFramework=\HalfMark,
    BuildProcess=\FullMark
  }
  \MidRule
  \Row{prop={\bfseries Use Cases}}
  \Row{prop={Decoy Request},
    DeploymentAgent=\EmptyMark,
    KubernetesOperator=\EmptyMark,
    SecondaryContainers=\HalfMark,
    LifecycleHooks=\EmptyMark,
    Honeypots=\EmptyMark,
    ReverseProxies=\FullMark,
    FileModifications=\HalfMark,
    RuntimeInstrumentation=\FullMark,
    LdPreload=\FullMark,
    Ptrace=\FullMark,
    EbpfMonitoring=\HalfMark,
    NetfilterRouting=\HalfMark,
    EbpfRouting=\HalfMark,
    FilesystemImplementations=\EmptyMark,
    VolumeMounts=\EmptyMark,
    PolicyEngine=\EmptyMark,
    Cni=\FullMark,
    DeceptionFramework=\FullMark,
    BuildProcess=\HalfMark
  }
  \Row{prop={Fake API},
    DeploymentAgent=\HalfMark,
    KubernetesOperator=\EmptyMark,
    SecondaryContainers=\HalfMark,
    LifecycleHooks=\EmptyMark,
    Honeypots=\HalfMark,
    ReverseProxies=\FullMark,
    FileModifications=\HalfMark,
    RuntimeInstrumentation=\FullMark,
    LdPreload=\FullMark,
    Ptrace=\FullMark,
    EbpfMonitoring=\HalfMark,
    NetfilterRouting=\HalfMark,
    EbpfRouting=\HalfMark,
    FilesystemImplementations=\EmptyMark,
    VolumeMounts=\EmptyMark,
    PolicyEngine=\EmptyMark,
    Cni=\FullMark,
    DeceptionFramework=\FullMark,
    BuildProcess=\HalfMark
  }
  \Row{prop={Honeytoken},
    DeploymentAgent=\FullMark,
    KubernetesOperator=\FullMark,
    SecondaryContainers=\FullMark,
    LifecycleHooks=\FullMark,
    Honeypots=\EmptyMark,
    ReverseProxies=\EmptyMark,
    FileModifications=\FullMark,
    RuntimeInstrumentation=\HalfMark,
    LdPreload=\HalfMark,
    Ptrace=\HalfMark,
    EbpfMonitoring=\HalfMark,
    NetfilterRouting=\EmptyMark,
    EbpfRouting=\EmptyMark,
    VolumeMounts=\FullMark,
    FilesystemImplementations=\FullMark,
    PolicyEngine=\FullMark,
    Cni=\EmptyMark,
    DeceptionFramework=\FullMark,
    BuildProcess=\FullMark
  }
  \MidRule
  \Row{prop={\bfseries Topological Properties}}
  \Row{prop={Deployment Mode~\cite{Han2018:DeceptionTechniquesComputer}},
    DeploymentAgent=\DeplAddedTo,
    KubernetesOperator=\DeplAddedTo,
    SecondaryContainers=\DeplAddedTo,
    LifecycleHooks=\DeplAddedTo,
    Honeypots=\hspace{0.425em}\DeplAddedTo\tnote{1},
    ReverseProxies=\DeplInFrontOf,
    FileModifications=\DeplAddedTo,
    RuntimeInstrumentation=\DeplInFrontOf,
    LdPreload=\DeplInFrontOf,
    Ptrace=\DeplInFrontOf,
    EbpfMonitoring=\DeplAddedTo,
    NetfilterRouting=\DeplInFrontOf,
    EbpfRouting=\DeplInFrontOf,
    FilesystemImplementations=\DeplInFrontOf,
    VolumeMounts=\DeplAddedTo,
    PolicyEngine=\DeplAddedTo,
    Cni=\DeplInFrontOf,
    DeceptionFramework=\DeplBuiltIn,
    BuildProcess=\DeplBuiltIn
  }
  \Row{prop={Layer of Decept.~\cite{Han2018:DeceptionTechniquesComputer}},
    DeploymentAgent=\LayerApplication\tnote{2},
    KubernetesOperator=\LayerSystem,
    SecondaryContainers=\LayerApplication,
    LifecycleHooks=\LayerApplication,
    Honeypots=\LayerNetwork,
    ReverseProxies=\LayerApplication,
    FileModifications=\LayerData\tnote{3},
    RuntimeInstrumentation=\LayerApplication,
    LdPreload=\LayerApplication,
    Ptrace=\LayerApplication,
    EbpfMonitoring=\LayerSystem,
    NetfilterRouting=\LayerSystem,
    EbpfRouting=\LayerSystem,
    FilesystemImplementations=\LayerData\tnote{2},
    VolumeMounts=\LayerData\tnote{3},
    PolicyEngine=\LayerSystem,
    Cni=\LayerNetwork,
    DeceptionFramework=\LayerApplication,
    BuildProcess=\LayerData\tnote{3},
  }
  \Row{prop={\enskip Application Dim.},
    DeploymentAgent=\DimContainer,
    KubernetesOperator=\NanMark,
    SecondaryContainers=\DimContainer,
    LifecycleHooks=\DimContainer,
    Honeypots=\NanMark,
    ReverseProxies=\DimContainer,
    FileModifications=\DimContainer,
    RuntimeInstrumentation=\DimRuntime,
    LdPreload=\DimLibraries,
    Ptrace=\DimContainer,
    EbpfMonitoring=\DimKernel,
    NetfilterRouting=\DimKernel,
    EbpfRouting=\DimKernel,
    FilesystemImplementations=\DimKernel,
    VolumeMounts=\DimContainer,
    PolicyEngine=\NanMark,
    Cni=\NanMark,
    DeceptionFramework=\DimSource,
    BuildProcess=\DimContainer,
  }
  \Row{prop={K8s Control Plane},
    DeploymentAgent=\NoMark,
    KubernetesOperator=\YesMark,
    SecondaryContainers=\NoMark,
    LifecycleHooks=\NoMark,
    Honeypots=\NoMark,
    ReverseProxies=\NoMark,
    FileModifications=\NoMark,
    RuntimeInstrumentation=\NoMark,
    LdPreload=\NoMark,
    Ptrace=\NoMark,
    EbpfMonitoring=\NoMark,
    NetfilterRouting=\NoMark,
    EbpfRouting=\NoMark,
    FilesystemImplementations=\NoMark,
    VolumeMounts=\NoMark,
    PolicyEngine=\YesMark,
    Cni=\YesMark,
    DeceptionFramework=\NanMark,
    BuildProcess=\NanMark,
  }
  \MidRule
  \Row{prop={\bfseries Technical Properties}}
  \Row{prop={Payload Modification},
    DeploymentAgent=\YesMark,
    KubernetesOperator=\YesMark,
    SecondaryContainers=\YesMark,
    LifecycleHooks=\YesMark,
    Honeypots=\NoMark,
    ReverseProxies=\YesMark,
    FileModifications=\YesMark,
    RuntimeInstrumentation=\YesMark,
    LdPreload=\YesMark,
    Ptrace=\YesMark,
    EbpfMonitoring=\NoMark,
    NetfilterRouting=\NoMark,
    EbpfRouting=\NoMark,
    FilesystemImplementations=\YesMark,
    VolumeMounts=\YesMark,
    PolicyEngine=\YesMark,
    Cni=\YesMark,
    DeceptionFramework=\YesMark,
    BuildProcess=\YesMark,
  }
  \Row{prop={\enskip Indirect Data Access},
    DeploymentAgent=\NoMark,
    KubernetesOperator=\NoMark,
    SecondaryContainers=\NoMark,
    LifecycleHooks=\NoMark,
    Honeypots=\NanMark,
    ReverseProxies=\YesMark,
    FileModifications=\NoMark,
    RuntimeInstrumentation=\YesMark,
    LdPreload=\YesMark,
    Ptrace=\YesMark,
    EbpfMonitoring=\NanMark,
    NetfilterRouting=\NanMark,
    EbpfRouting=\NanMark,
    FilesystemImplementations=\YesMark,
    VolumeMounts=\NoMark,
    PolicyEngine=\hspace{0.425em}\NoMark\tnote{4},
    Cni=\YesMark,
    DeceptionFramework=\NoMark,
    BuildProcess=\NoMark,
  }
  \Row{prop={Zero Downtime}}
  \Row{prop={\enskip ... for intial setup},
    DeploymentAgent=\YesMark,
    KubernetesOperator=\YesMark,
    SecondaryContainers=\hspace{0.425em}\YesMark\tnote{5},
    LifecycleHooks=\NoMark,
    Honeypots=\YesMark,
    ReverseProxies=\NoMark,
    FileModifications=\hspace{0.425em}\YesMark\tnote{6},
    RuntimeInstrumentation=\hspace{0.425em}\YesMark\tnote{7},
    LdPreload=\NoMark,
    Ptrace=\YesMark,
    EbpfMonitoring=\YesMark,
    NetfilterRouting=\YesMark,
    EbpfRouting=\YesMark,
    FilesystemImplementations=\NoMark,
    VolumeMounts=\NoMark,
    PolicyEngine=\YesMark,
    Cni=\NoMark,
    DeceptionFramework=\NoMark,
    BuildProcess=\NoMark,
  }
  \Row{prop={\enskip ... for live changes},
    DeploymentAgent=\YesMark,
    KubernetesOperator=\YesMark,
    SecondaryContainers=\hspace{0.425em}\YesMark\tnote{5},
    LifecycleHooks=\NoMark,
    Honeypots=\YesMark,
    ReverseProxies=\YesMark,
    FileModifications=\hspace{0.425em}\YesMark\tnote{6},
    RuntimeInstrumentation=\YesMark,
    LdPreload=\YesMark,
    Ptrace=\YesMark,
    EbpfMonitoring=\YesMark,
    NetfilterRouting=\YesMark,
    EbpfRouting=\YesMark,
    FilesystemImplementations=\YesMark,
    VolumeMounts=\YesMark,
    PolicyEngine=\YesMark,
    Cni=\YesMark,
    DeceptionFramework=\YesMark,
    BuildProcess=\NoMark,
  }
  \Row{prop={Software Dependencies},
    DeploymentAgent=\NothingMark,
    KubernetesOperator=\ResKubernetes,
    SecondaryContainers=\ResKubernetes,
    LifecycleHooks=\ResOciCompliant,
    Honeypots=\NothingMark,
    ReverseProxies=\ResServiceMesh,
    FileModifications=\NothingMark,
    RuntimeInstrumentation=\ResManagedRuntime,
    LdPreload=\ResLibc,
    Ptrace=\NothingMark,
    EbpfMonitoring=\ResEbpf,
    NetfilterRouting=\ResNetfilter,
    EbpfRouting=\ResEbpf,
    FilesystemImplementations=\NothingMark,
    VolumeMounts=\ResOciCompliant,
    PolicyEngine=\ResPolicyEngine,
    Cni=\ResCniPlugin,
    DeceptionFramework=\NothingMark,
    BuildProcess=\NothingMark,
  }
  \MidRule
  \Row{prop={\bfseries Operational Props.}}
  \Row{prop={Detectability~\cite{Bowen2009:BaitingAttackersUsing}},
    DeploymentAgent=\GoodMark,
    KubernetesOperator=\FairMark,
    SecondaryContainers=\GoodMark,
    LifecycleHooks=\PoorMark,
    Honeypots=\ExcellentMark,
    ReverseProxies=\ExcellentMark,
    FileModifications=\PoorMark,
    RuntimeInstrumentation=\GoodMark,
    LdPreload=\FairMark,
    Ptrace=\FairMark,
    EbpfMonitoring=\ExcellentMark,
    NetfilterRouting=\PoorMark,
    EbpfRouting=\ExcellentMark,
    FilesystemImplementations=\GoodMark,
    VolumeMounts=\PoorMark,
    PolicyEngine=\GoodMark,
    Cni=\GoodMark,
    DeceptionFramework=\ExcellentMark,
    BuildProcess=\GoodMark
  }
  \Row{prop={Simplicity},
    DeploymentAgent=\FairMark,
    KubernetesOperator=\FairMark,
    SecondaryContainers=\GoodMark,
    LifecycleHooks=\ExcellentMark,
    Honeypots=\GoodMark,
    ReverseProxies=\GoodMark,
    FileModifications=\GoodMark,
    RuntimeInstrumentation=\PoorMark,
    LdPreload=\PoorMark,
    Ptrace=\PoorMark,
    EbpfMonitoring=\FairMark,
    NetfilterRouting=\FairMark,
    EbpfRouting=\FairMark,
    FilesystemImplementations=\FairMark,
    VolumeMounts=\GoodMark,
    PolicyEngine=\GoodMark,
    Cni=\PoorMark,
    DeceptionFramework=\PoorMark,
    BuildProcess=\FairMark,
  }
  \Row{prop={Maintainability},
    DeploymentAgent=\PoorMark,
    KubernetesOperator=\PoorMark,
    SecondaryContainers=\FairMark,
    LifecycleHooks=\FairMark,
    Honeypots=\ExcellentMark,
    ReverseProxies=\GoodMark,
    FileModifications=\GoodMark,
    RuntimeInstrumentation=\GoodMark,
    LdPreload=\GoodMark,
    Ptrace=\GoodMark,
    EbpfMonitoring=\GoodMark,
    NetfilterRouting=\GoodMark,
    EbpfRouting=\GoodMark,
    FilesystemImplementations=\GoodMark,
    VolumeMounts=\GoodMark,
    PolicyEngine=\GoodMark,
    Cni=\GoodMark,
    DeceptionFramework=\FairMark,
    BuildProcess=\FairMark,
  }
  \Row{prop={Scalability},
    DeploymentAgent=\FairMark,
    KubernetesOperator=\ExcellentMark,
    SecondaryContainers=\GoodMark,
    LifecycleHooks=\GoodMark,
    Honeypots=\GoodMark,
    ReverseProxies=\ExcellentMark,
    FileModifications=\FairMark,
    RuntimeInstrumentation=\GoodMark,
    LdPreload=\GoodMark,
    Ptrace=\GoodMark,
    EbpfMonitoring=\ExcellentMark,
    NetfilterRouting=\GoodMark,
    EbpfRouting=\GoodMark,
    FilesystemImplementations=\GoodMark,
    VolumeMounts=\ExcellentMark,
    PolicyEngine=\ExcellentMark,
    Cni=\ExcellentMark,
    DeceptionFramework=\GoodMark,
    BuildProcess=\GoodMark,
  }
  \MidRule
  \Row{prop={\bfseries Properties of Efficacy}}
  \Row{prop={Inconspicuousness~\cite{Bowen2009:BaitingAttackersUsing}},
    DeploymentAgent=\GoodMark,
    KubernetesOperator=\GoodMark,
    SecondaryContainers=\GoodMark,
    LifecycleHooks=\GoodMark,
    Honeypots=\PoorMark,
    ReverseProxies=\FairMark,
    FileModifications=\ExcellentMark,
    RuntimeInstrumentation=\ExcellentMark,
    LdPreload=\ExcellentMark,
    Ptrace=\ExcellentMark,
    EbpfMonitoring=\ExcellentMark,
    NetfilterRouting=\GoodMark,
    EbpfRouting=\GoodMark,
    FilesystemImplementations=\GoodMark,
    VolumeMounts=\GoodMark,
    PolicyEngine=\FairMark,
    Cni=\ExcellentMark,
    DeceptionFramework=\ExcellentMark,
    BuildProcess=\ExcellentMark,
  }
  \Row{prop={Non-interference~\cite{Bowen2009:BaitingAttackersUsing}},
    DeploymentAgent=\FairMark,
    KubernetesOperator=\FairMark,
    SecondaryContainers=\FairMark,
    LifecycleHooks=\FairMark,
    Honeypots=\ExcellentMark,
    ReverseProxies=\PoorMark,
    FileModifications=\PoorMark,
    RuntimeInstrumentation=\BadMark,
    LdPreload=\BadMark,
    Ptrace=\BadMark,
    EbpfMonitoring=\GoodMark,
    NetfilterRouting=\FairMark,
    EbpfRouting=\FairMark,
    FilesystemImplementations=\FairMark,
    VolumeMounts=\FairMark,
    PolicyEngine=\PoorMark,
    Cni=\PoorMark,
    DeceptionFramework=\BadMark,
    BuildProcess=\PoorMark,
  }
\end{KeyValTable}

%% file: content/06-discussion.tex
\section{Discussion}
\label{sec:discussion}

We compare technical methods, 
show opportunities for future research,
and note challenges that we encountered. 

\textbf{Technical methods at the container- and pod-level can be difficult to detect for adversaries,
  but also risk interfering with the genuine flow of an application.}
Bringing cyber deception techniques closer to the application layer
is supposed to increase their effectiveness in engaging adversaries without them noticing%
~\cite{Han2017:EvaluationDeceptionBasedWeb}.
Invasive container- and pod-specific methods~(\S\ref{sec:container-pod-methods})
cover a wide range of technical capabilities%
~\cite{Kern2024:InjectingSharedLibraries,Araujo2020:ImprovingCybersecurityHygiene},
while remaining stealthy and thus difficult for an adversary to detect.
Compared to \mReverseProxies{}~\tReverseProxies{} and \mHoneypots{}~\tHoneypots{},
which can often be fingerprinted~\citehoneypotfingerprinting{},
we expect application-level hooks and patches to be harder to detect.
Stealthy technical methods paired with enticing payloads%
~\cite{%
  Sahin2022:ApproachGenerateRealistic,
  Sahin2020:LessonsLearnedSunDEW
}
leave little room for countermeasures by adversaries.
Methods with a low performance overhead may even evade timing-based fingerprinting%
\cite{Mukkamala2007:DetectionVirtualEnvironments},
especially in cloud environments, which exhibit high network latencies%
~\cite{Kahlhofer2023:BenchmarkingFunctionHook}.
Extensive studies on the fingerprintability
of these methods have not yet appeared in the literature.

These invasive methods share many of the properties and capabilities
of methods that would require developer interaction,
but they may introduce more application inconsistencies, incompatibilities, and vulnerabilities%
\cite{%
  Islam2021:CHIMERAAutonomousPlanning,
  Ramaswamy2010:KatanaHotPatching,
  Araujo2020:ImprovingCybersecurityHygiene}.
%
%
Thus, it is reasonable that reverse proxies receive most of the attention instead%
~\citereverseproxysolutions{}.
However, since this pattern often comes with a significant performance overhead%
~\citesidecaroverheadinresearch{},
we believe that future research is needed to reduce this overhead,
e.g., with hybrid solutions that offload work to the kernel,
inspired by industry trends~\citesidecaroverheadinindustry{}.




\textbf{We encourage future research on
  invasive technical methods that are also non-interfering, maintainable, and scalable;
  on leveraging cloud-native platforms; and on software frameworks for deception.}
%
Our review showed that some methods that leverage Kubernetes design patterns%
~\cite{Ibryam2019:KubernetesPatterns}
(\mKubernetesOperator{}~\tKubernetesOperator{},
\mSecondaryContainers{}~\tSecondaryContainers{},
\mLifecycleHooks{}~\tLifecycleHooks{},
and \mVolumeMounts{}~\tVolumeMounts{})
score relatively high 
in non-interference and scalability;
\mVolumeMounts{}~\tVolumeMounts{} are also low-maintenance.
These methods have been used for logging and monitoring,
also in the context of cyber deception%
~\cite{Gupta2023:HoneyKubeDesigningDeploying,Gupta2021:HoneyKubeDesigningHoneypot},
but they do not yet appear to be used for placing decoys.


Future research on the usability of deception methods for system operators
also seems interesting: What can be done to make deception policies transparent and
easy to integrate, scale, and maintain?
Custom Resource Definitions (CRDs)~\cite{Ibryam2019:KubernetesPatterns}, in Kubernetes,
could be used to transparently define and manage deception policies. 

Although few studies limit themselves to methods
without developer interaction (as we do), 
we are surprised that there is no work yet on software frameworks (or plugin systems)
that make it easy to embed cyber deception at the design phase.
Certainly, the wide variety of software frameworks that one would like to support
will be a challenge~\cite{Sahin2020:LessonsLearnedSunDEW},
but this problem has been solved in other domains:
OpenTelemetry~\cite{TheOpenTelemetryAuthors:OpenTelemetry}, in the observability domain,
succeeded in establishing a standard for collecting
(custom) telemetry data from applications.


\textbf{The best technical methods for cyber deception must also to be paired
  with enticing payloads, which in turn can be mined by technical methods.}
This work did not research which payloads
(e.g., filenames and their contents) are most enticing to adversaries%
~\cite{%
  Sahin2022:ApproachGenerateRealistic,
  Sahin2022:MeasuringDevelopersWeb,
  Sahin2020:LessonsLearnedSunDEW,
  Bercovitch2011:HoneyGenAutomatedHoneytokens,
  Bowen2010:AutomatingInjectionBelievable
}.
Some of the methods we explored in this work can also provide
data for precisely such experiments, e.g. by using \mEbpfMonitoring{}~\tEbpfMonitoring{}
to monitor which files in a file system show conspicuous access patterns,
or by detecting API endpoints of running applications with \mReverseProxies{}~\tReverseProxies{}.
The latter could then be used to automatically derive new deceptive API endpoints from that,
as demonstrated by \citeauthor{Sahin2022:ApproachGenerateRealistic}%
~\cite{Sahin2022:ApproachGenerateRealistic}.



\textbf{Technical methods should not be considered in isolation, but in combination.}
We have chosen not to map the interdependencies between the methods,
as there are many creative combinations.
Our work should not be read with the intention of picking the ``best'' method,
but with the motivation to find a suitable combination of methods.
As an example, suppose one wants to automate
the placement of honeytokens in all container file systems.
In addition, we want to automatically receive alerts on access attempts to these files.
We could use a \mPolicyEngine{}~\tPolicyEngine{}
such as Kyverno~\cite{TheKyvernoAuthors:Kyverno} and specify two policies.
The first policy defines \mVolumeMounts{}~\tVolumeMounts{}
that place a honeytoken (the decoy) in every container file system at \texttt{/var/run/secrets/},
which is the path where Kubernetes secrets are typically mounted.
The second policy will add an \mEbpfMonitoring{}~\tEbpfMonitoring{}-based
tracing policy (the captor) with Tetragon~\cite{TheTetragonAuthors:Tetragon}
which monitors for file access attempts to this path and sends alerts to operators.
A \mKubernetesOperator{}~\tKubernetesOperator{} could deploy and manage such policies.


\textbf{Cloud-native environments and various technical circumstances
  made it difficult to clearly categorize methods.}
The distinction of the layer of deception~\cite{Han2018:DeceptionTechniquesComputer} into
network, system, application, and data needed further definition
to better fit the world of clusters, platforms, pods, and containers~(\S\ref{sec:properties}).

Some technical properties were difficult to assign; e.g.,
whether installing a \mRuntimeInstrumentation{}~\tRuntimeInstrumentation{} agent
at runtime is possible or not depends not only
on the specific runtime environment and its version,
but also on machine architectures and the scope of the agent to be installed.
Since many (modern) methods are also changing rapidly,
it may be necessary to re-evaluate some of them.



%% file: content/07-conclusion.tex
\section{Conclusion}
\label{sec:conclusion}

This work reviewed \numMethods{}~technical methods to achieve application layer cyber deception.
We identified properties for evaluating these methods
and categorized them into container- and pod-specific,
kernel- and network-specific, and container platform-specific methods.
We separately group supporting methods,
and commonly used methods such as self-contained honeypots and reverse proxies.

This overview shall help practitioners and researchers
to bring models from game theory and creative deception ideas
into contact with production-grade systems,
which require robust, scalable, and maintainable technical solutions.
It is particularly challenging when methods need to be added after the design phase,
but still need to be close to the application layer, as this not only increases complexity,
but also the risk of interfering with genuine application assets.
However, being less conspicuous to adversaries
and more independent of the software development lifecycle
keeps this approach interesting.
Our work also finds less explored methods that seem to manage this balancing act:
Leveraging design patterns of cloud-native platforms
and building software frameworks for cyber deception.

%% file: content/08-appendix.tex
\section{Excluded Properties}
\label{sec:appendix:excluded-properties}

Many more properties have been introduced to categorize cyber deception techniques%
~\cite{%
  Han2018:DeceptionTechniquesComputer,
  Fan2018:EnablingAnatomicView,
  Nawrocki2016:SurveyHoneypotSoftware},
some of which are still considered ``difficult to formalize and measure''%
~\cite{Han2018:DeceptionTechniquesComputer}.
We briefly explain which ones we excluded:
\begin{itemize}
  \item The \emph{fidelity or interaction level}%
        ~\cite{Mokube2007:HoneypotsConceptsApproaches,Boumkheld2019:HoneypotTypeSelection}
        (low, medium, high) describes how ``immersive'' a honeypot is for an adversary.
        Technical methods do not limit the creative freedom here, 
        especially not when a combination of methods is used.
  \item We exclude properties of decoys such as
        \emph{believability}, \emph{enticingness}, \emph{conspicuousness}
        and \emph{variability}, as introduced by
        \citeauthor{Bowen2009:BaitingAttackersUsing}~\cite{Bowen2009:BaitingAttackersUsing},
        because they are typically influenced by concrete names and payloads,
        but not by technical properties.
  \item We exclude typical measures for honeypots that lie outside the application layer%
        ~\cite{Fan2018:EnablingAnatomicView,Nawrocki2016:SurveyHoneypotSoftware}
        such as their \emph{role} (server, client), \emph{physicality or virtuality},
        and also their \emph{deployment strategy}~\cite{Scottberg2002:InternetHoneypotsProtection}.
  \item We exclude measures for captors, as introduced by
        \citeauthor{Fan2018:EnablingAnatomicView}~\cite{Fan2018:EnablingAnatomicView},
        e.g., describing attack \emph{monitoring}, \emph{prevention},
        \emph{detection}, \emph{response}, and \emph{profiling}.
        We see concrete implementations on these independent of general technical methods.
\end{itemize}

\section{Guide for Qualitative Evaluation of Methods}
\label{sec:appendix:rubric}


\textbf{D-C-S Type.}
From a system operator's perspective, we categorize if a type fits a method's
\emph{primary use case} perfectly~(\FullMark), partially~(\HalfMark), or not at all~(\EmptyMark).
A partial fit indicates that a method is technically able to represent that type,
although, it would be uncommon or unreasonable to use a method like that.

\begin{itemize}
  \item \textbf{Decoy.} Does this method \emph{typically} represent
        -- not necessarily create (like a supporting method) --
        the entity with which an adversary will interact with?
        (e.g., honeypot, honeytoken, reverse proxy) 
  \item \textbf{Captor.} Is this method \emph{typically} used to
        \emph{detect} attacks or events on decoys?
        (e.g., by observing logs, monitoring events)
  \item \textbf{Support.} Does this method \emph{typically} assist in
        \emph{deploying, implementing, or configuring} another method?
        (e.g., the process that places decoys)
\end{itemize}


\textbf{Supported Use Cases.}
Categorized based on whether that method \emph{is capable}~(\FullMark)
or \emph{not capable}~(\EmptyMark) to realize
the use cases that we introduced in \S\ref{sec:problem-statement}, regardless of complexity.
Reliance on other (supporting) methods (e.g., user space processes)
that might restrict the applicability of the method when considered in isolation,
shall be indicated with \emph{partial capability}~(\HalfMark) then.

\begin{itemize}
  \item \textbf{Decoy request.}
        Can this method \emph{append deceptive payloads} to the HTML response of an application?
  \item \textbf{Fake API.}
        Can this method \emph{respond to a request for ``/admin/api''} on behalf of the application?
  \item \textbf{Honeytoken.}
        Can this method \emph{create or represent} a ``service-token'' file in a file system?
\end{itemize}


\textbf{Deployment Mode.}
We adopt the categorization from
\citeauthor{Han2018:DeceptionTechniquesComputer}~\cite{Han2018:DeceptionTechniquesComputer}
from a system operator's perspective.

\begin{itemize}
  \item \DeplBuiltIn~\textbf{Built-in.}
        Does this method add something at \emph{design phase}?
        (e.g., source code, build process).
  \item \DeplInFrontOf~\textbf{In-front of.}
        Does this method \emph{interpose} something between an adversary and a genuine application?
        (e.g., proxies, filters, hooks)
  \item \DeplAddedTo~\textbf{Added-to.}
        Does this method include, integrate, or modify something \emph{at runtime}?
        (e.g., add file to file system, add pod to cluster, change setting)
  \item \DeplStandAlone~\textbf{Stand-alone.}
        Is this method \emph{isolated} from the target system?
        (e.g., honey accounts on the internet, honeypots outside your Kubernetes cluster)
\end{itemize}

\textbf{Layer of Deception.}
We adopt the categorization from
\citeauthor{Han2018:DeceptionTechniquesComputer}~\cite{Han2018:DeceptionTechniquesComputer}
and interpret it for cloud-native architectures such as Kubernetes.
If the layer of deception is \emph{application},
or if it affects applications (e.g., kernel-based techniques that affect applications),
we also assign it to one of the finer levels,
exactly as described in \S\ref{par:topological-properties} under topological properties.
\begin{itemize}
  \item \textbf{Network.}
        Does this method realize deception techniques that are
        \emph{''accessible over the network
          and that are not bound to any specific host configuration''}%
        ~\cite{Han2018:DeceptionTechniquesComputer}?
        (e.g., typical honeypots, or methods that produce false network topologies)
  \item \textbf{System.}
        Does this method realize deception techniques that are \emph{bound to hosts}?
        In Kubernetes, we consider nodes, and also the whole platform part of this layer.
        (e.g., kernel-based techniques, or platform-wide policy management)
  \item \textbf{Application.}
        Does this method realize deception techniques \emph{bound to specific applications}?
        We consider containers, pods and their file system part of this layer.
        (e.g., honeytokens in file systems, runtime instrumentation techniques)
  \item \textbf{Data.}
        Does this method realize deception techniques
        by \emph{leveraging data, documents, or files}? (e.g., honeytokens in a file system)
\end{itemize}

\textbf{Kubernetes Plane.}

\begin{itemize}
  \item \textbf{Data plane.}
        Does this method \emph{primarily} act on the data plane
        (on the node, pod, or container level)?
  \item \textbf{Control plane.}
        Does this method primarily \emph{act, work with, configure, or need} the control plane?
\end{itemize}


\textbf{Payload Modification.}
Is this method capable of
\emph{modifying, extending, or removing data payloads on disk, or in transit},
directly or indirectly, but definitely on its own?
More specifically, we refer to \emph{data payloads}, not just (packet) header fields or metadata.
If so, we further ask if payloads \emph{can} -- but not must -- be modified indirectly,
i.e., \emph{by changing it on read, in transit, through a proxy or wrapper},
but never by directly editing the original data item in the file system, database, or source code.

\textbf{Zero Downtime.}
Can this method be first installed without requiring an \emph{application restart}?
Once installed, \emph{can} the deception be easily
\emph{enabled and disabled, modified, and also reconfigured}?


\textbf{Detectability.}
How capable is this method to \emph{send alerts, log events, or monitor honeypots}?
We only consider the capability that fits the primary use case of that method, e.g.,
a method that places honeytokens in the file system
will be evaluated on its ability to detect access attempts to them,
and we will not consider its detection capabilities for other use cases.
We assess detectability by focusing on how much control we have over
the piece of code or component that detects or monitors:

\begin{itemize}
  \item \ExcellentMark~Do we have \emph{direct control} over this code?
  \item \GoodMark~Would we \emph{need to integrate} this code or component into an existing system?
  \item \FairMark~Would we need to integrate this code or component into an existing system
        \emph{and rely} on more than one other software component?
  \item \PoorMark~Would we need another \emph{supporting method}
        or have \emph{little control} over a persistent solution?
  \item \BadMark~Is this \emph{not possible} at all?
\end{itemize}

\textbf{Simplicity.}
On a technical level, how simple -- not easy -- is this method to implement or configure?

\begin{itemize}
  \item \ExcellentMark~Does this method require only a \emph{small} amount
        of code or configuration to be added or changed?
  \item \GoodMark~Is this method similar in technical complexity
        to \emph{typical} work tasks of an engineer or operator?
  \item \FairMark~Will this method take a typical engineer
        \emph{a few weeks} to implement?
  \item \PoorMark~Could this method easily require
        a \emph{full engineering team} (min. 3-7 developers) to implement?
\end{itemize}

\textbf{Maintainability.}
From a system operator's and deception designer's perspective,
how much maintenance does this method require?

\begin{itemize}
  \item \ExcellentMark~Does this require almost \emph{no maintenance}?
  \item \GoodMark~Must this only be checked \emph{a few times per year}?
  \item \FairMark~Must this be checked \emph{every few days or weeks}?
  \item \PoorMark~Could this method easily require
        a \emph{full engineering team} (min. 3-7 developers) to maintain?
\end{itemize}

\textbf{Scalability.}
From a system operator's perspective,
how easy is it to scale this method to hundreds an thousands of computer systems?

\begin{itemize}
  \item \ExcellentMark~Is this method cluster-wide
        almost \emph{by default}?
  \item \GoodMark~Does this method require only
        a \emph{small} amount of code or configuration to be scaled cluster-wide?
  \item \FairMark~Will this method take a typical engineer
        \emph{a few weeks} of work before it can be scaled cluster-wide?
  \item \PoorMark~Could this method easily require
        a \emph{full engineering team} (min. 3-7 developers) to scale it?
  \item \BadMark~Is it almost \emph{impossible} to scale this method?
\end{itemize}


\textbf{Inconspicuousness.}
From an adversary's perspective, how obvious or clearly visible is that deception method?

\begin{itemize}
  \item \ExcellentMark~Is this method \emph{very difficult or impossible} to detect
        by observing the behavior of an application,
        even if the adversary has compromised the container hosting the application?
  \item \GoodMark~Is this method \emph{very difficult or impossible} to detect
        by observing the behavior of an application
        \emph{unless} the adversary has compromised the container hosting the application?
  \item \FairMark~If the adversary spends a \emph{few weeks} working on this,
        will this method eventually be detected?
  \item \PoorMark~Is this method fairly \emph{easy} to detect, even for an average adversary?
  \item \BadMark~Is this method so \emph{obvious}
        that it doesn't even take a good hacker to discover it?
\end{itemize}

\textbf{Non-interference.}
From a system operator's perspective,
how low is the risk of this method interfering with genuine application and system assets?

\begin{itemize}
  \item \ExcellentMark~Is this method almost
        \emph{isolated, or stand-alone} from the real production system?
  \item \GoodMark~Does this method only \emph{passively monitor}?
  \item \FairMark~Does this method \emph{add processes, files, or configuration}
        with little risk of interference?
  \item \PoorMark~Does this method \emph{instrument}
        genuine applications at runtime or \emph{modify} data?
  \item \BadMark~Could this method easily \emph{crash} an application?
\end{itemize}